\documentclass[10pt,journal,compsoc]{IEEEtran}

\ifCLASSOPTIONcompsoc
  \usepackage[nocompress]{cite}
\else
  \usepackage{cite}
\fi

\ifCLASSINFOpdf
  \usepackage[pdftex]{graphicx}
  \graphicspath{{./figs/}}
  \DeclareGraphicsExtensions{.pdf,.jpeg,.png}
\else
   \usepackage[dvips]{graphicx}
   \graphicspath{{./figs/}}
   \DeclareGraphicsExtensions{.eps}
\fi

\usepackage{amsmath}
\usepackage{algorithmic}
\usepackage{array}
\usepackage{url}

\graphicspath{{figs/}} 

\usepackage{amssymb}
\usepackage{amsmath} 
\usepackage{amsfonts}

\usepackage{hyperref}

\newcommand {\mm}[1] {\ifmmode{#1}\else{\mbox{\(#1\)}}\fi}
\newcommand{\Rspace}        {\mm{\mathbb{R}}}
\newcommand{\Xspace}        {\mm{\mathbb{X}}}

\newcommand{\etal}{\textit{et al.}}
\newcommand{\mywlog}{\textit{w.l.o.g.}}
\newcommand{\cf}{\textit{cf.}}

\newcommand{\IM}        {\mm{M}}
\newcommand{\IV}        {\mm{V}}

\newcommand{\wrt}        {\mm{\textit{w.r.t.\,}}}

\newcommand{\LCA}       {\mm{lca}}

\newcommand{\T}        {\mm{\mathcal{T}}}

\newcommand{\para}[1]        {\vspace{1pt}\noindent{\textbf{#1}}}
\newcommand{\denselist}{\vspace{-3pt} \itemsep -2pt\parsep=-1pt\partopsep -2pt}

\usepackage{amsthm} 
\usepackage{amsfonts}
\usepackage{mathtools}

\theoremstyle{definition}
\newtheorem{definition}{Definition}[section]

\usepackage{amsthm}

\usepackage{subfigure}

\newcommand{\footref}[1]{\textsuperscript{\ref{#1}}}

\newcommand{\MG}        {\mm{\mathsf{Moving\,Gaussian}}}
\newcommand{\CF}  {\mm{\mathsf{Corner\,Flow}}}
\newcommand{\HF}        {\mm{\mathsf{Heated\,Flow}}}
\newcommand{\VS}        {\mm{\mathsf{Vortex\,Street}}}
\newcommand{\RS}        {\mm{\mathsf{Red\,Sea}}}
\newcommand{\Wing}        {\mm{\mathsf{Wing}}}

\newcommand{\ie} {{\textit{i.e.}}}

\usepackage{mathptmx}
\DeclareMathAlphabet{\mathcal}{OMS}{cmsy}{m}{n}
\usepackage{mathtools}
\usepackage{float}

\newcommand{\grad}[1]     {{\nabla {#1}}}

\begin{document}

\title{Geometry-Aware Merge Tree Comparisons for \\ Time-Varying Data with  Interleaving Distances}

\author{Lin Yan, Talha Bin Masood, Farhan Rasheed, Ingrid Hotz, Bei Wang
\IEEEcompsocitemizethanks{\IEEEcompsocthanksitem 
Lin Yan and Bei Wang are with the University of Utah. 
E-mails: lynne.h.yan@gmail.com, beiwang@sci.utah.edu.
\IEEEcompsocthanksitem Talha Bin Masood, Farhan Rasheed, and Ingrid Hotz are  with Link\"{o}ping University. 
E-mails: talha.bin.masood@liu.se, farhan.rasheed@liu.se, ingrid.hotz@liu.se.}}

\IEEEtitleabstractindextext{
\begin{abstract}
Merge trees, a type of topological descriptors, serve to identify and summarize the topological characteristics associated with scalar fields. 
They present a great potential for the analysis and visualization of time-varying data.  
First, they give compressed and topology-preserving representations of data instances. 
Second, their comparisons provide a basis for studying the relations among data instances, such as their distributions, clusters, outliers, and periodicities. 
A number of comparative measures have been developed for merge trees. 
However, these measures are often computationally expensive since they implicitly consider all possible correspondences between critical points of the merge trees. 
In this paper, we perform geometry-aware comparisons of merge trees using labeled interleaving distances. 
The main idea is to decouple the computation of a comparative measure into two steps: a \emph{labeling} step that generates a correspondence between the critical points of two merge trees, and a \emph{comparison} step that computes distances between a pair of labeled merge trees by encoding them as matrices. 
We show that our approach is general, computationally efficient, and practically useful. 
Our general framework makes it possible to integrate geometric  information of the data domain in the labeling process. 
At the same time, it reduces the computational complexity since not all possible correspondences have to be considered. 
We demonstrate via experiments that such geometry-aware merge tree comparisons help to detect \emph{transitions}, \emph{clusters}, and \emph{periodicities} of time-varying datasets, as well as to \emph{diagnose} and \emph{highlight} the topological changes between adjacent data instances.

\end{abstract}

\begin{IEEEkeywords}
Merge trees, merge tree metrics, topological data analysis, topology in visualization
\end{IEEEkeywords}
}

\maketitle

\IEEEdisplaynontitleabstractindextext
\IEEEpeerreviewmaketitle


\section{Introduction}
\label{sec:introduction}

The efficient analysis of large and complex data is an essential component of a modern scientific workflow. 
The key ingredients in data analysis are the identification, tracking, and comparison of features expressing essential structures in the data.  
To this end, {\em topological data analysis} (TDA) has proven to provide fundamental tools for visual data analysis in terms of abstraction and  summarization. 
Topological descriptors for scalar field data, such as persistence diagrams, barcodes, merge trees, contour trees, Reeb graphs, and Morse--Smale complexes, are among the most widely used applied topological tools in visualization. 
These descriptors present a great potential for the analysis and visualization of time-varying data.  
First, they give compressed and topology-preserving representations of data instances. 
Second, their comparisons provide a basis for studying the relations among data instances, such as their distributions, clusters, outliers, and periodicities; see the work of Yan \etal~\cite{YanMasoodSridharamurthy2021} for a recent survey.  
		
In this paper, we are interested in merge trees, which are topological descriptors that record the connectivity among the sublevel sets of scalar fields. 
Merge trees have seen many applications in science and engineering, including cyclone tracking~\cite{Valsangkar2018}, burning structure analysis~\cite{BremerWeberTierny2010}, and symmetry extraction in materials science~\cite{ThomasNatarajan2011,MasoodThomasNatarajan2013}, to name a few. 
To employ merge trees for time-varying data or ensembles, a key challenge is to choose an appropriate similarity or distance measure for their comparisons. 
A number of comparative measures have been developed for merge trees in the literature (see~\autoref{sec:related-work} and~\cite{YanMasoodSridharamurthy2021}), of which many effectively ``forget'' about the geometric information from the data domain in the comparative process. 

We take a different perspective to utilize merge trees in studying time-varying data, and ask the following question: How can we design a merge tree comparative measure that integrates geometric information from the data domain?  
We hypothesize that by enriching a merge tree with geometrical information, a comparative measure defined on such enriched merge trees will be sensitive to local or global geometry and thus become beneficial for real-world applications where such geometry is important. 

Our work is also motivated by the study of \emph{information content} within a topological descriptor. 
A few recent efforts have linked information theory with topology. 
Merelli \etal~\cite{MerelliRuccoSloot2015} introduced an entropic measure called the \emph{persistence entropy} that captures information within a barcode. 
Edelsbrunner \etal~\cite{EdelsbrunnerVirkWagner2019} studied TDA in information space by utilizing information-theoretic distances. 
Recent work by Brown \etal~\cite{BrownBobrowskiMunch2020}	 posed  questions regarding the information content of Reeb graphs, which are topological descriptors closely related to merge trees. 
Specifically, they asked the following questions: ``What information is encoded by a Reeb graph?" ``How much information can we recover about the original data from the Reeb graph by solving an inverse problem?''~\cite{BrownBobrowskiMunch2020}.   
We are motivated by similar questions regarding the merge tree: 

\begin{itemize}\denselist
\item How much geometric information could we add to a merge tree, so as to increase/enrich its information content, while at the same time improving the comparative process among a pair of 	``enriched" merge trees in real-world applications? 
\item Will geometry-aware comparative measures for a set of merge trees improve our understanding of the time-varying data, in terms of its distributions, clusters, and outliers?
\end{itemize}

 \begin{figure}[!ht]
    \centering
    \includegraphics[width=1.0\columnwidth]{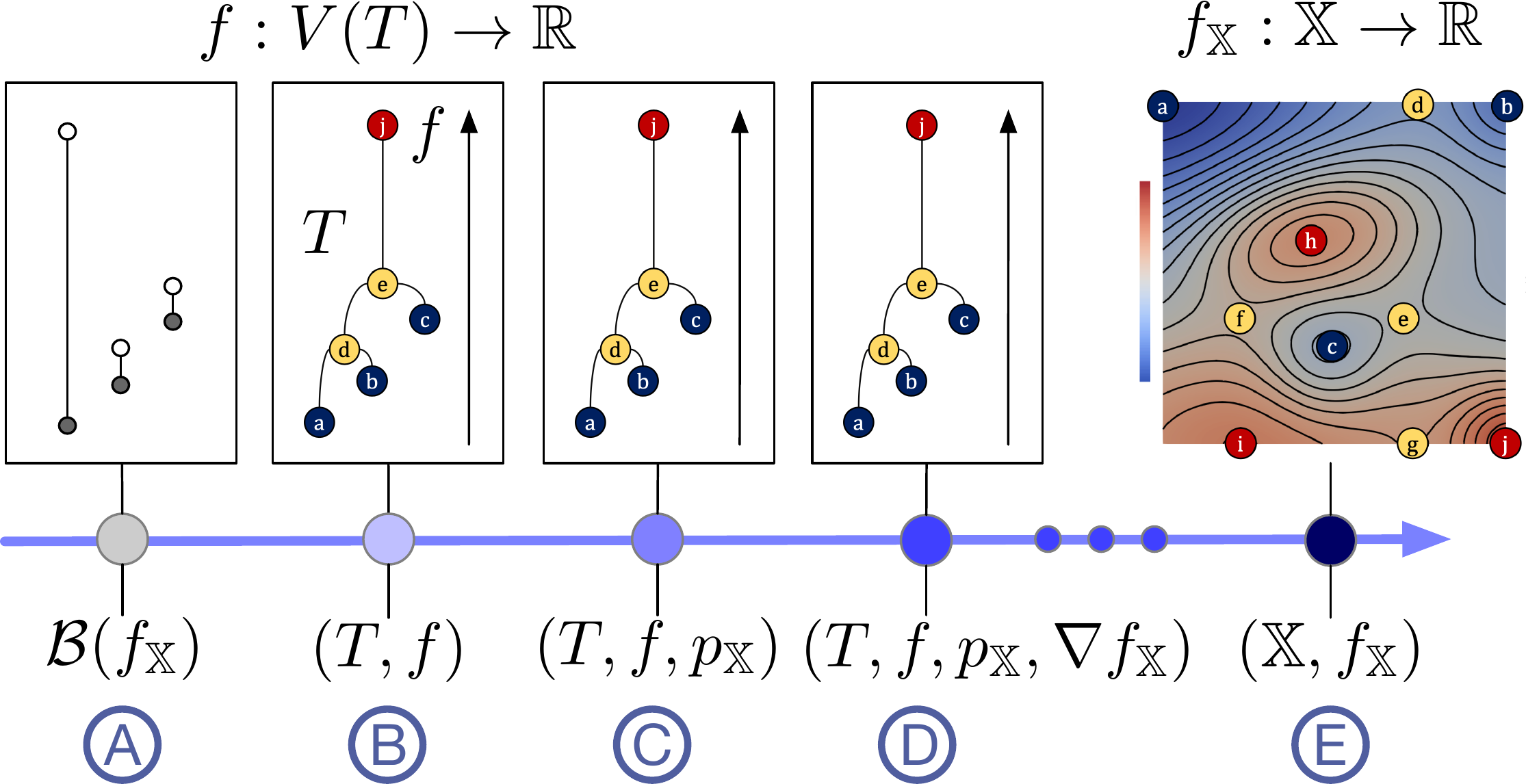}
    \vspace{-6mm}
    \caption{Given a scalar field $f_{\Xspace}: \Xspace \to \Rspace$, from left to right, we increase the geometric content of the topological descriptors associated with $f_{\Xspace}$: (A) the barcode of $f_{\Xspace}$; (B) the ``classic" merge tree $(T, f)$; (C) adding the coordinates of critical points $p_{\Xspace}$ of $f_{\Xspace}$; (D) adding the gradient information $\nabla
 f_{\Xspace}$ of $f_{\Xspace}$; (E) the original scalar field data.}
\label{fig:mt-information}
\end{figure}

\noindent We further illustrate the above thought process in~\autoref{fig:mt-information}. 
Formally, given a scalar field defined on a connected domain $f_{\Xspace}: \Xspace \to \Rspace$, a merge tree records the connectivity of its sublevel sets and is represented as a pair $(T, f)$, that is, a finite rooted tree $T$ equipped with a function defined on its vertices $f: V(T) \to \Rspace$; where $f$ is a restriction of $f_{\Xspace}$ to its critical points (see also~\autoref{sec:background}). 
As we move from left to right in~\autoref{fig:mt-information}, we increase the geometric content of a topological descriptor for $f_{\Xspace}$.  
We begin with the persistence barcode of $f_{\Xspace}$ in~\autoref{fig:mt-information}A. 
By adding slightly more geometric information, that is, how bars in the barcode are glued together, we obtain the ``classic'' merge tree of $f_{\Xspace}$, denoted as $(T, f)$ in~\autoref{fig:mt-information}B.  
The connection between barcodes and merge trees first appeared (rather implicitly) via the Elder Rule in~\cite[Page 150]{EdelsbrunnerHarer2010}, and was further explored in~\cite{Curry2018, CatanzaroCurryFasy2020, KanariGarinHess2020}. 
Although many existing comparative measures on merge trees rely on such a classic definition of merge trees, the focus of this paper is to develop comparative measures on merge trees that encode the geometry of the data domain. 
Specifically, we focus on enriching a merge tree by encoding the coordinates of critical points $p_\Xspace$ of $f_{\Xspace}$ in ~\autoref{fig:mt-information}C. 
It is also possible to add the gradient information $\nabla  f_{\Xspace}$ in~\autoref{fig:mt-information}D. 
Finally, a merge tree can be enriched with the scalar field $f_{\Xspace}$ itself, possibly rendering the tree redundant in~\autoref{fig:mt-information}E.

Our objective is to perform geometry-aware comparisons of merge trees. 
We explore a general notion of merge tree called the \emph{leaf  labeled merge tree}~\cite{GasparovicMunchOudot2019,YanWangMunch2020}, which is an abstract tree equipped with a scalar function and a labeling of its leaves.   
Our main idea is to \emph{decouple} the computation of a comparative metric for a pair of (enriched) merge trees into two steps: 
\begin{itemize}\denselist
\item[1.] A \emph{labeling} step that generates a correspondence between the critical points of two merge trees using various geometric information of the data domain;  
\item[2.] A \emph{comparison} step that computes distances between  pairs of labeled merge trees represented as matrices. 
\end{itemize} 
The labeling step makes it possible to integrate geometric information of the data domain, such as the locations of critical points and the gradients of the underlying scalar fields. 
It also allows the encoding of application-specific domain knowledge.
Furthermore, it reduces the computational complexity since not all possible correspondences have to be considered. 
We provide several heuristic strategies for labeling leaves in a merge tree, namely, \emph{tree mapping}, \emph{Euclidean mapping}, and their \emph{hybrid mapping}. 
We also discuss a \emph{Morse mapping} strategy based on the gradients of an input scalar field. 
For the comparison step, we use the \emph{labeled interleaving distance}~\cite{GasparovicMunchOudot2019}, which encodes the labeling information within matrix representations of merge trees.

Using datasets in scientific simulations, we experimentally evaluate and compare our framework against well-established comparative measures for merge trees, using the bottleneck distance and tree edit distance~\cite{SridharamurthyMasoodKamakshidasan2020} as the baseline. 
In summary: 
\begin{itemize}\denselist
\item We provide a general and unifying two-step framework that supports geometry-aware comparisons of merge trees for time-varying data; 
\item We demonstrate that our proposed framework can help detect  \emph{transitions}, \emph{clusters}, and \emph{periodicities} of a time-varying dataset, as well as to \emph{diagnose} and \emph{highlight} the topological changes between adjacent data instances.
\end{itemize}
Finally, our framework is open source at \url{https://github.com/tdavislab/MergeTreeMetric}.

\section{Related Work}
\label{sec:related-work}

\para{Topological descriptors for scalar fields.}
A \emph{topological descriptor}, in our context, serves to describe and identify the topological characteristics associated with a data instance (e.g., a scalar field, a vector field, a tensor field, or a multi-field). 
Most topological descriptors we work with originate from the Morse theory~\cite{Milnor1963}. 
Well-established topological descriptors for scalar fields include persistence diagrams~\cite{EdelsbrunnerLetscherZomorodian2002},  barcodes~\cite{CarlssonZomorodianCollins2004}, 
merge trees, contour trees~\cite{CarrSnoeyinkAxen2003}, Reeb graphs~\cite{Reeb1946} and their discretization, mapper graphs~\cite{SinghMemoliCarlsson2007}, and Morse--Smale complexes~\cite{EdelsbrunnerHarerZomorodian2003}.  
      
Additional descriptors/visual representations exist that integrate topological \emph{and} geometric information of a scalar field~\cite{YanMasoodSridharamurthy2021}. 
Correa \etal~\cite{CorreaBremerLindstrom2011} introduced extremum graphs for the design of topological spines, which consist of edges between extrema and shared saddles of scalar fields. 
Extrema graphs were shown to capture the geometric proximity of the extrema better than the contour trees~\cite{CorreaBremerLindstrom2011}. 
Thomas and Natarajan~\cite{ThomasNatarajan2013} further introduced an augmented extremum graph for geometry-aware symmetry detection in scalar fields. 
Saikia \etal~\cite{SaikiaSeidelWeinkauf2014} introduced the extended branch decomposition graph, which consists of the branch decompositions of all subtrees of the merge tree. 
Narayanan \etal~\cite{NarayananThomasNatarajan2015} further introduced a complete extremum graph that associates proximity information for all pairs of extrema. 

\para{Merge trees and applications.}
In this paper, we focus on a type of topological descriptor called merge trees, which track the evolution of the connected components of the sublevel sets of scalar fields.
Merge trees have been used to identify and track cyclones~\cite{Valsangkar2018} and bubbles in Raleigh-Taylor instabilities~\cite{LaneyBremerMascarenhas2007}, as well as to analyze burning cells~\cite{BremerWeberTierny2010} from combustion simulations.
Contour trees~/~merge trees also show up in feature tracking~\cite{ReininghausKastenWeinkauf2012, ReininghausKotavaGunther2011, SaikiaWeinkauf2017, SolerPlainchaultConche2018}. 

Merge trees can also be used to study vector and tensor fields.   
Wang \etal~\cite{WangRosenSkraba2013} used the merge tree to extract robust critical points from vector fields. 
Wang and Hotz used the merge tree of an anisotropy field to establish the foundation for a stability measure of degenerate points in the tensor field~\cite{WangHotz2017}. 
Jankowai \etal~\cite{JankowaiWangHotz2019} utilized such a measure for robust extraction and simplification of 2D tensor field topology.   

\para{Topological metrics and kernels.}
We explore scalar fields by comparing their corresponding topological descriptors, which requires a measure of similarity or dissimilarity between them.
The main challenge for the practical use of any similarity measure is its \emph{stability} and \emph{computability}. 
Recently, existing metrics and kernels for topological descriptors have been investigated extensively~\cite{YanMasoodSridharamurthy2021}.
 
For persistence diagrams, well-established metrics include the bottleneck distance~\cite{Cohen-SteinerEdelsbrunnerHarer2007,EdelsbrunnerHarer2008} and the Wasserstein distance~\cite{Cohen-SteinerEdelsbrunnerHarer2010}. 
They are often used as a comparison baseline for the distance between other types of descriptors (such as merge trees).
However, these distances do not incorporate information regarding relations (e.g.,~nesting) between (sub-)level sets, which are often necessary for analysis~\cite{BeketayevYeliussizovMorozov2014}. 
Besides, persistence diagrams do not have the structure of an inner product space ({\ie}~Hilbert space), making it difficult to interface them with machine learning.
Instead, several topological kernels have been proposed in the literature~\cite{LeYamada2018,ReininghausHuberBauer2015}, making persistence diagrams suitable for learning tasks such as kernel support vector machines.

For Reeb graphs and their variances, namely, merge trees and contour trees, many metrics have been introduced (e.g.~\cite{MorozovBeketayevWeber2013,BauerGeWang2014,CarriereOudot2017,BauerLandiMemoli2020,SridharamurthyMasoodKamakshidasan2020}. 
However, many of these metrics remain theoretical and do not have practical implementations. 
Beketayev \etal~\cite{BeketayevYeliussizovMorozov2014} used the branch decompositions of merge trees to define a distance between them. 
Their algorithm constructed all possible matchings among pairs of branch decompositions and selected the minimum cost matching among them. 
Several distances draw inspirations from the Gromov-Hausdorff distance for measuring metric distortion~\cite{BauerGeWang2014,MemoliSmithWan2019}.
Bauer \etal~\cite{BauerGeWang2014} introduced the functional distortion distance between the Reeb graphs, which was then utilized to compare metric graphs~\cite{DeyShiWang2015}. 
Sridharamurthy \etal~\cite{SridharamurthyMasoodKamakshidasan2020} introduced an edit distance between merge trees that admits efficient computation. 
Their edit distance is defined as the minimum cost of a set of restricted edit operations (e.g.,~delete, insert, and relabel) that transforms one merge tree into another.

Apart from global structural comparison, distance metrics have also been developed based upon local structures. 
Thomas and Natarajan~\cite{ThomasNatarajan2011} defined a similarity measure for comparing subtrees of a contour tree and use it to group similar subtrees together to extract symmetric structures in scalar fields.
Saikia \etal~\cite{SaikiaSeidelWeinkauf2014} performed structural comparison and extracted repeating topological structures by comparing all subtrees of two merge trees against each other. 
For hybrid descriptors, Narayanan \etal~\cite{NarayananThomasNatarajan2015} introduced a feature-aware distance metric between extremum graphs that is based on the maximum common subgraph of the complete extremum graphs.   

The works of Gasparovic \etal~\cite{GasparovicMunchOudot2019} and Yan \etal~\cite{YanWangMunch2020} are most relevant to the current paper. 
Gasparovic \etal~\cite{GasparovicMunchOudot2019} introduced an easily computable metric called the \emph{labeled interleaving distance} that can be used to compare labeled merge trees.
Yan \etal~\cite{YanWangMunch2020} adapted such a distance in practice for computing average merge trees and visualizing uncertainty. 
They introduced a few heuristic strategies that generate  correspondences between merge trees, which set the foundation for the labeling step in our framework.  
Compared with~\cite{YanWangMunch2020}, we perform a  more systematic comparative study on how geometry aware comparative measures for merge trees improve our understanding of time-varying data under various visualization tasks, including the detection of transitions, clusters, and periodicities. 
We further introduce time-varying pivot tree and dummy vertices  strategies, which are shown to be more effective in study the time-varying data, compared with the global pivot tree and dummy leaves strategies first introduced by Yan \etal~\cite{YanWangMunch2020}.

\section{Background}
\label{sec:background}
In this section, we review the necessary background on scalar field topology surrounding the notions of merge trees, labeled merge trees, and distances between the labeled merge trees. 

\subsection{Merge Trees and Their Variants}
\begin{figure}[!ht]
    \centering
    \includegraphics[width=.99\columnwidth]{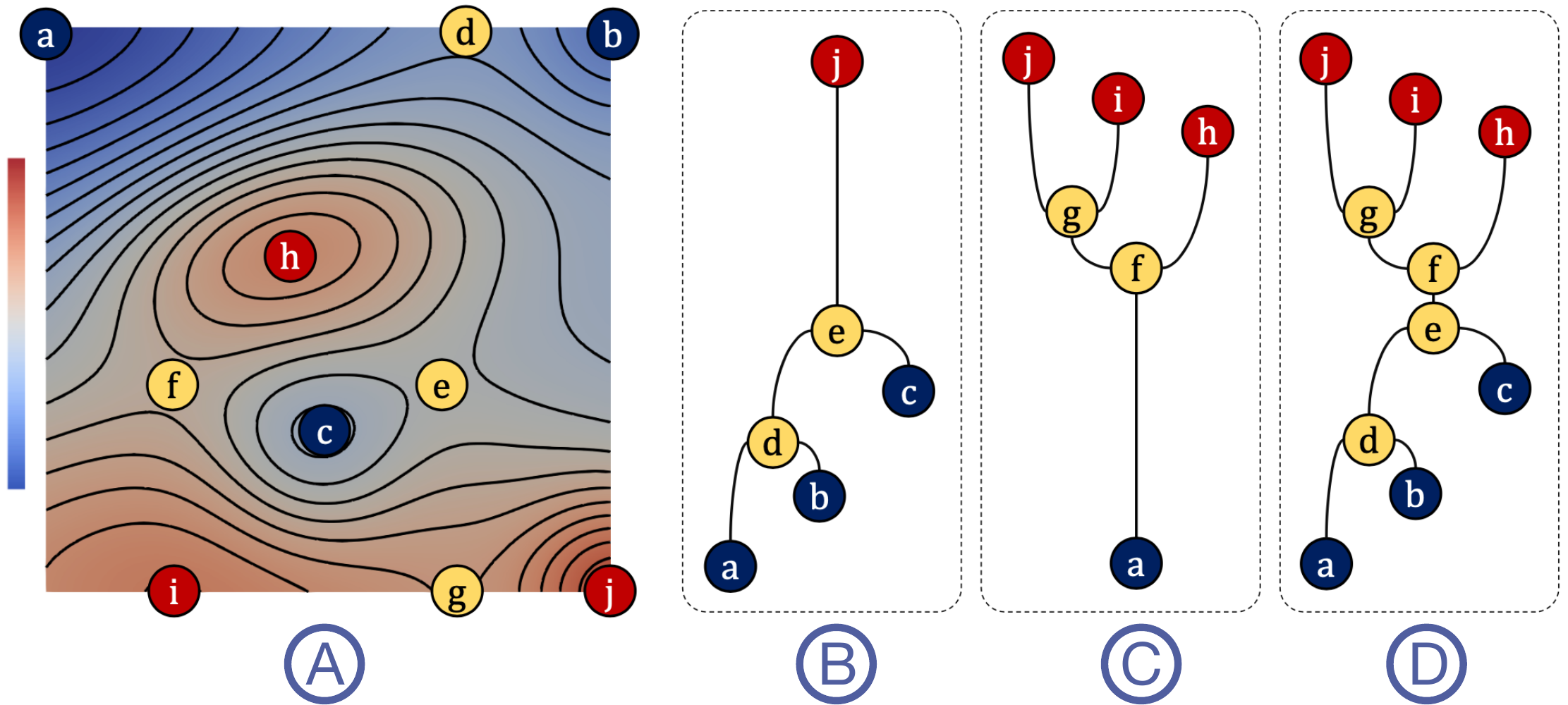}
    \vspace{-2mm}
    \caption{For the scalar field $f$ in (A): the merge tree of $f$ (\ie, the join tree), the merge tree of $-f$ (\ie, the split tree), and the contour tree of $f$ are illustrated in (B), (C), and (D), respectively.} 
\label{fig:mt-example}
\end{figure}

\para{Scalar-field-induced merge trees.}
Given a scalar field $f: \Xspace \to \Rspace$ defined on a connected  domain $\Xspace$, a (scalar field induced) merge tree records the connectivity of its sublevel sets. 
Two points $x, y \in \Xspace$ are considered \emph{equivalent} \wrt  $f$, $x \sim y$, if they have the same function value, that is, $f(x) = f(y) = a$, and if they belong to the same connected component of the sublevel set $\Xspace_a:= f^{-1}(-\infty, a]$, for some $a \in \Rspace$.
A \emph{merge tree} is the quotient space $\Xspace/{\sim}$ obtained by gluing together points in $\Xspace$ that are equivalent under the relation $\sim$. 
Intuitively, it keeps track of the evolution of connected  components in $\Xspace_a$ as $a$ increases; see~\autoref{fig:mt-example}B for an example. 
Specifically, leaves ({\ie},~non-root vertices with degree 1) in a merge tree  represent the creation of a component at a local minimum, internal vertices (of degree $\geq 3$) represent the merging of components, and the root (a degree 1 vertex) represents the entire space as a single component. 

Throughout this paper, we denote our \emph{data} of interest as a pair $(\Xspace, f)$, that is, a connected topological space $\Xspace$ together with a scalar field $f:\Xspace \to \Rspace$.
The quotient space $\Xspace/{\sim}$ is a new topological space that effectively ``forgets" about certain information regarding the data $(\Xspace,f)$ such as the locations of the critical points and the gradient of $f$. 
 
\para{Labeled merge trees.}
In this paper, we work with a more general notion of merge trees defined below, which can be considered as an abstract tree coupled with a function $f$ on its vertices, while being oblivious of the possible existence of a scalar field that gives rise to the merge tree. 

\begin{definition}
\label{def:mt}
A \emph{merge tree} is a pair $(T,f)$ of a finite rooted tree $T$ with vertex set $V(T)$ and a function $f:V(T) \to \Rspace \cup \{\infty\}$ such that (i) adjacent vertices do not have equal function value, (ii) every non-root vertex has exactly one neighbor with higher function value, and (iii) the root is the only vertex with the value $\infty$~\cite[Def. 2.1]{GasparovicMunchOudot2019}.
\end{definition}
\noindent Def.~\ref{def:mt} is closely related to that of \emph{treegrams}~\cite{SmithChowdhuryMemoli2016}, which is a certain generalization of a dendrogram~\cite{CarlssonMemoli2010}.
Specifically, we focus on the notion of a labeled merge tree. 
Let $[n]$ denote a set of labels $\{1,2,\cdots,n\}$.
\begin{definition}
\label{def:lmt}
A \emph{labeled merge tree}, denoted as a triple $\T=(T, f, \pi)$, consists of a merge tree $(T,f)$ together with a labeling $\pi:[n] \to V(T)$ that is surjective on the set of leaves~\cite[Def. 2.2]{GasparovicMunchOudot2019}. 
\end{definition}
\noindent $\pi$ is not required to be injective; thus, a vertex can have multiple labels. 
$\pi$ also allows labels for non-leaves by thinking of these as degenerate labeled leaves~\cite{GasparovicMunchOudot2019}.
For this paper, we work with \emph{leaf-labeled merge trees}, where $L(T) \subset V(T)$ represents its leave set.  
Unless otherwise specified, they are referred to as labeled merge trees for the remainder of the paper. 
We study the space of labeled merge trees that share the \emph{same} label set $[n]$.

\para{Join and split trees.} 
For convenience, we refer to the merge tree of $f$ as the \emph{join} tree and the merge tree of $-f$ as the \emph{split} tree (following the convention    in~\cite{CarrSnoeyinkAxen2003}).   
A join tree (\autoref{fig:mt-example}B) tracks the connected component of the sublevel sets of $f$ while a split tree (\autoref{fig:mt-example}C) tracks that of the superlevel sets of $f$.
The leaves of a join tree are local minima of $f$, while the leaves of a split tree are local maxima of $f$. 
Combining a join tree and a split tree of $f$ gives rise to its \emph{contour tree}~\cite{CarrSnoeyinkAxen2003} (\autoref{fig:mt-example}D), which captures the connectivity among level sets. 
As illustrated in~\autoref{sec:results}, studying the join tree or the split tree of time-varying data offers different perspectives  on their topological signatures. 

\para{Tree and Euclidean distances between vertices.}
Given an unlabeled merge tree $(T, f)$, the \emph{intrinsic tree distance} $d_t$ between pairs of vertices is induced by $f: V(T) \to \Rspace$. 
For any pair of vertices $x, y \in V(T)$, it is defined as 
$$d_t(x,y)=|f(x)-f(\LCA(x,y))| + |f(y)-f(\LCA(x,y))|,$$ 
where $\LCA(x,y)$ denotes the lowest common ancestor of $x$ and $y$ in $T$. 
In other words, $d_t(x,y)$ captures the shortest path between $x$ and $y$ measured by function value differences to their lowest common ancestor. 

On the other hand, let $\tau_i: |T^i| \rightarrow \Rspace^2$ denote the geometric embeddings of $T_i$ into the spatial domain (for $i=1,2$).   
The \emph{Euclidean distance $d_e$} between a pair of vertices $x, y \in V(T)$ is induced by its geometric embedding $\tau$. 
It is defined as 
$$d_e(x, y) = || \tau(x)-\tau(y)||_2.$$ 

\subsection{Interleaving Distances Between Merge Trees}

A number of metrics may be defined on the space of labeled merge trees. 
 In fact, any metric defined on unlabeled merge trees may be extended to labeled ones by forgetting the label information; which likely turns a metric into a pseudometric~\cite{GasparovicMunchOudot2019}. 

For a labeled merge tree, again let $\LCA(u,v)$ denote the \emph{lowest common ancestor} of a pair of vertices. We have $\LCA(u,u) = u$. 
Let $f(\LCA(u,v))$ denote its function value.

\begin{figure}[!ht]
    \centering
    \includegraphics[width=0.86\columnwidth]{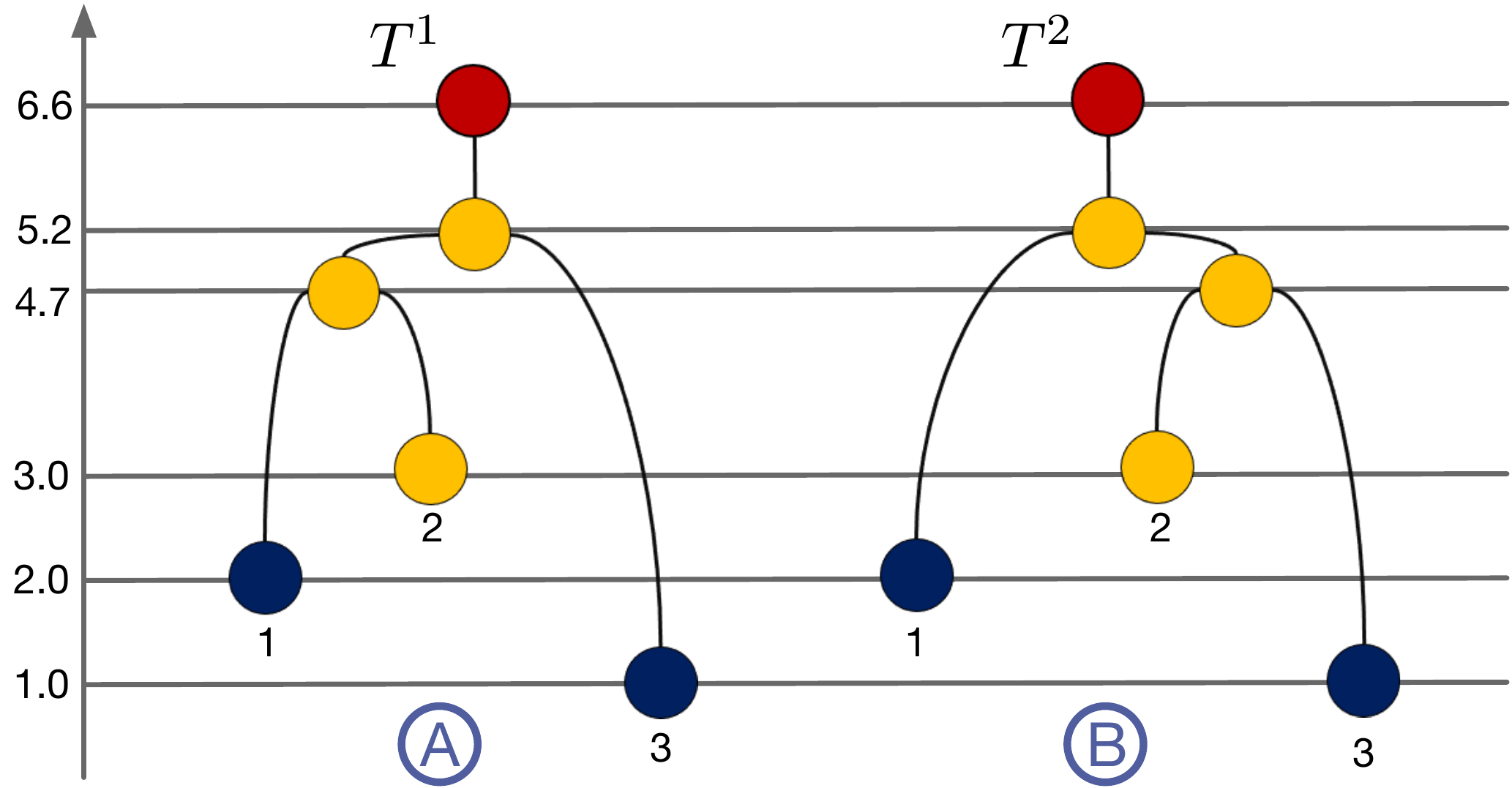}
    \vspace{-2mm}
    \caption{Labeled merge trees $T^1$ and $T^2$ with a shared label set $[3]:=\{1, 2, 3\}$.} 
\label{fig:induced-matrices}
\end{figure}

The \emph{induced matrix} of a labeled merge tree $\T=(T, f, \pi)$, denoted as $\IM(T,f,\pi)$, is the matrix $\IM_{ij} = f( \LCA(\pi(i),  \pi(j)))$~\cite[Def. 2.6]{GasparovicMunchOudot2019}.
As shown in \autoref{fig:induced-matrices}A-B, the induced matrices of labeled merge trees $T^1$ and $T^2$ with a shared label set $[3]:=\{1,2,3\}$ are defined as follows:
\[
M^1 =
\begin{bmatrix}
2.0 & 4.7 &5.2 \\
\cdot & 3.0 & 5.2 \\
\cdot & \cdot & 1.0
\end{bmatrix},\;
M^2 =
\begin{bmatrix}
2.0 & 5.2 &5.2 \\
\cdot & 3.0 & 4.7 \\
\cdot & \cdot & 1.0
\end{bmatrix}.
\]

$\IM \in \Rspace^{n \times n}$ is symmetric; therefore, it is common to store $\IM$ as an upper triangular matrix with $n(n+1)/2$ nonzero entries. 
We use such a convention in our paper. 
The nonzero terms in $\IM$ can be linearized into a vector of length $n(n+1)/2$, called the \emph{cophenetic vector} of $\T$~\cite{CardonaMirRossello2013}, denoted as $\IV$.

Turning a labeled merge tree into a matrix (or a vector) enables us to use distances between matrices (or vectors) to obtain distances between trees. 
We work with the \emph{cophenetic metrics} first introduced by Cardona \etal~\cite{CardonaMirRossello2013} for phylogenetic trees.

\begin{definition}
Given two labeled merge trees $\T^1=(T_1, f_1, \pi_1)$ and $\T^2=(T_2, f_2, \pi_2)$ that share the same set of labels $[n]$, the $L^{\infty}$-, $L^{1}$-, and $L^{2}$-cophenetic metrics are defined, respectively, as distances between their corresponding induced matrices $\IM^1$ and $\IM^2$~\cite{CardonaMirRossello2013}:    
$$d^{\infty}(\T^1, \T^2) = ||\IM^1 - \IM^2||_{\infty},$$
$$d^1(\T^1, \T^2) = ||\IM^1 - \IM^2||_{1},$$
$$d^2(\T^1, \T^2) = ||\IM^1 - \IM^2||_{2}.$$
\label{def:distance}
\end{definition}
\noindent$||M||_{\infty}$, $||M||_{1}$ and $||M||_{2}$ denote the vector norms, that is, $||M||_{\infty}=\max_{ij} |M_{ij}|$, $||M||_1=\sum_{ij}|M_{ij}|$ and $||M||_2=\sqrt{\sum_{ij}|M_{ij}|^2}$. 
In other words, the above distances correspond to the $L^{\infty}$, $L^{1}$, and $L^{2}$ distances between the cophenetic vectors of $\T^1$ and $\T^2$ respectively. 
It is important to point out that $d^{\infty}$, $d^{1}$, and $d^2$ work only on labeled merge trees since we need the (same set of) labels to have a well-defined induced matrix. 

$d^{\infty}$ is also referred to as the \emph{labeled interleaving distance}, denoted as $d_I$, between a pair of labeled merge trees~\cite[Def. 2.13]{GasparovicMunchOudot2019} (and subsequently~\cite[Def. 3.3]{YanWangMunch2020}), due to its close connection to the interleaving distance of persistence modules~\cite{ChazalCohen-SteinerGlisse2009,ChazalSilvaGlisse2016}.
Such a connection is described explicitly in~\cite{MunchStefanou2018}. 
The interleaving distance of persistence modules has been adapted to merge trees~\cite{MorozovBeketayevWeber2013,TouliWang2019}, Reeb graphs~\cite{SilvaMunchPatel2016,Curry2014}, and Reeb spaces~\cite{MunchWang2016}, via a category-theoretic perspective~\cite{BubenikSilvaScott2014, SilvaMunchStefanou2018}. 
$d_I$ has also been used to compare single-linkage hierarchical clustering~\cite{SmithChowdhuryMemoli2016}. 
Recent work has established both theoretical~\cite{GasparovicMunchOudot2019} and algorithmic~\cite{GasparovicMunchOudot2019,YanWangMunch2020} foundations for the space of merge trees under the interleaving distances, including computing a form of structural averages for uncertainty visualization~\cite{YanWangMunch2020}. 
For the remainder of this paper, we work with the interleaving distance $d_I$ with the understanding that $d_I := d^{\infty}$ for labeled merge trees. 
For example, as illustrated in \autoref{fig:induced-matrices}, the interleaving distance $d_I(\T^1, \T^2) = ||\IM^1 - \IM^2||_{\infty} = 0.5$. 
As illustrated in \autoref{sec:results}, one of the main advantages of the labeled interleaving distance $d_I$ is that, as an $L^\infty$ type distance, it helps to diagnose which entries in the induced matrices are likely responsible for the given distance. 

\subsection{Other Distances}
We review a few other distance metrics that are applicable for labeled merge trees. Recall any metric defined on unlabeled merge trees is applicable by ignoring the labeling. 

\para{Bottleneck and 1-Wasserstein distance.} 
Given two persistence diagrams $X^1$, $X^2$ and a bijection $\eta: X^1 \to X^2$, the \emph{bottleneck distance} between $X^1$ and $X^2$ is defined to be 
\begin{align}
W_{\infty}(X^1, X^2) =  \adjustlimits \inf_{\eta: X^1 \to X^2}  \sup_{x \in X^1} ||x \in \eta(x)||_\infty.
\end{align} 
The \emph{q-Wasserstein distance} is
\begin{align}
W_{q}(X^1, X^2) = \left[ \inf_{\eta: X^1 \to X^2}  \sum_{x \in X^1} ||x \in \eta(x)||^q_\infty \right]^{1/q}.
\end{align} 
By definition, $W_q$ becomes $W_\infty$ by setting $q=\infty$. 
In dimension zero, which is the concern of this paper, there is a correspondence between a merge tree $T$ of $f$ and the persistence diagram $X$ of its sublevel set filtration. 
Specifically, the branch decomposition of $T$ of $f$ gives rise to the points in the persistence diagram of $f$. 
More generally, given a merge tree $(T,f)$ in the sense of Def.~\ref{def:mt}, we could directly define the \emph{persistence diagram of the merge tree}, denoted as $X_{T}$, by treating $T$ as a topological space and computing the 0-dimensional persistence of its sublevel set filtration, following a similar construction in~\cite{BauerGeWang2014}. 
Therefore, we could define the bottleneck distance between the merge trees as the bottleneck distance between the persistence diagrams of the merge trees, that is, 
\begin{align}
d_{B}(\T^1, \T^2) =  W_{\infty}(X_T^1, X_T^2).
\end{align}
Similarity, we work with $1$-Wasserstein distance between the merge trees, 
\begin{align}
d_{W}(\T^1, \T^2) =  W_{1}(X_T^1, X_T^2).
\end{align}

\para{Edit distance between merge tree.} 
The \emph{edit distance} between merge trees~\cite{SridharamurthyMasoodKamakshidasan2020} is defined as 
\begin{align}
\label{equation:edit}
d_E(\T^1,\T^2) = \min_S \{\gamma(S)\},
\end{align} 
where $S$ is a tree edit operation sequence from $T^1$ to $T^2$ that includes edit operations such as ``relabel", ``delete", and ``insert"; and $\gamma$ is a cost function that assigns a non-negative real number to each operation. 
$d_E$ is an adaptation of and a significant improvement on the constrained unordered tree edit distance~\cite{Zhang1996}.

\para{Distance between scalar fields.}
Finally, given a pair of scalar fields $f_1$ and $f_2$ -- each sampled at $N$ locations forming a vector of length $N$, denoted as $v_1$ and $v_2$ -- we work with Euclidean distances between them in the comparative study, defined as
$$d_{F}(f_1, f_2) = ||v_1 - v_2||_{2}.$$

\section{Geometry-Aware Merge Tree Metrics}
\label{sec:method}

The merge tree metrics described in~\autoref{sec:related-work} are unaware of the geometric information regarding the data domain,  such as the locations of the critical points or the gradient of the scalar field. 
These metrics are defined over the space of \emph{unlabeled merge trees}. 
To define a geometry-aware metric, we consider a more general space that allows for encoding such information. 
A suitable space is the space of \emph{labeled merge trees} introduced in~\autoref{sec:background}. 
Our two-step framework for merge tree comparisons includes: 
\begin{itemize}\denselist
\item[1.] \emph{Labeling}. This step generates a correspondence between the critical points of two merge trees by encoding the geometric information of the data domain. We describe several labeling strategies --  namely, \emph{tree mapping},  \emph{Euclidean mapping}, and \emph{hybrid mapping} -- based on topological and geometric information of the scalar fields, respectively. 
\item[2.] \emph{Comparison}. This step transforms labeled  merge trees into their \emph{induced matrices}, and computes distances between labeled merge trees by computing distances between their corresponding induced matrices. 
\end{itemize} 
In this section, we describe the labeling step in detail. 
Suppose we are given a time-varying dataset consisting of $l$ instances of scalar fields defined over a common domain with their corresponding merge trees. 
There are two requirements to transfer a set of unlabeled merge trees into labeled ones.  
First, we need to assign a common set of labels between leaves of merge trees, that is, to find leaf correspondences that capture properties of the data domain; see~\autoref{sec:mapping}.  
Second, we need to ensure the induced matrices of labeled merge trees are comparable. We introduce \emph{dummy nodes} and \emph{dummy leaves} to the trees to ensure that the resulting induced matrices are of the same size. The latter (dummy leaves) strategy was first introduced in~\cite{YanWangMunch2020}; however, we perform a more systematic study of both strategies in~\autoref{sec:dummy}.   
Finally, we introduce the notion of a \emph{time-varying pivot tree}, which is shown to be effective to capture topological transitions in a time-varying setting; see~\autoref{sec:pivot-tree}.

\subsection{Labeling Strategies}
\label{sec:mapping} 

Our leaf labeling strategies take as input a pair of unlabeled merge trees $T^1$ and $T^2$ that arise from a pair of 2D scalar fields $f_1$ and $f_2$. 
We consider a tree mapping strategy, a Euclidean mapping strategy,  and a hybrid mapping strategy, where the hybrid mapping strategy generalizes the other two strategies. 

Given an unlabeled merge tree $(T, f)$, recall that the intrinsic tree distance between a pair of vertices $x$ and $y$ in the tree is denoted by $d_t$, and the Euclidean distance between their geometric embeddings is denoted by $d_e$.  
$d_t$ and $d_e$ capture the topological and geometric relation between $x$ and $y$, respectively. 
To achieve a balance between topology and geometry, we define a hybrid distance as 
\[
d_h(x,y) = \lambda \cdot d_t + (1 - \lambda) \cdot d_e, 
\]
for $0 \leq \lambda \leq 1$. 
$d_h$ generalizes both $d_t$ and $d_e$. For $\lambda=1$, $d_h = d_t$ and for $\lambda = 0$, $d_h = d_e$.  
We thus focus on describing the hybrid mapping strategy, which finds a minimum cost matching between leaves based on their similarities among their $d_h$ distances to other vertices in the tree. 

\para{Initial label assignment.}
Given a pair of unlabeled merge trees $T^1$ and $T^2$, let $V(T^1)$ and $V(T^2$) denote their respective vertex sets, and $L(T^1)$ and $L(T^2)$ their leaf sets. 
The goal of an initial label assignment between $L(T^1)$ and $L(T^2)$ is to utilize topology, geometry, or prior knowledge of the data to establish initial correspondences (labels) between subsets of the leaves that share high similarities.   
These initially matched labels serve as ``anchors'' during the labeling process, where distances to these anchors are used to assess topological and geometric similarities among the remaining unlabeled leaves. 

We use geometric proximities of critical points in the domain under the Euclidean distance as an example. 
To define a shared label set, the tree with a larger number of leaves is chosen as the \emph{pivot tree} $T_p$.
The leaves of $T_p$ give rise to a \emph{pivot label set}, denoted $S_p:=[n]$, where $n = |L(T_p)|$.
Let $\pi_p: S_p \to L(T_p)$ denote a labeling of the pivot tree; 
{\mywlog}, assume $T_p = T^1$ and $\pi_1$ is its leaf labeling.   
This label set is assigned to $L(T^2)$ via $\pi_2: S_p \to L(T^2)$ with a minimum weight matching described below.

To initialize a labeling of $L(T^2)$, we assign a subset of labels in $S_p$ to $L(T^2)$ based on Euclidean distances between $L(T^2)$ and $L(T^1)$ in the embeddings. 
To do so, we construct a weighted, complete bipartite graph between $L(T^1)$ and $L(T^2)$ where the weight $w_{xy}$ between $x \in L(T^1)$ and $y \in L(T^2)$ is their Euclidean distance in the embedded space constrained by a non-negative threshold $\epsilon$:  $w_{xy} = d_e(\tau_1(x), \tau_2(y))$ if $d_e(\tau_1(x), \tau_2(y)) \leq \epsilon$; otherwise $w_{xy} = \infty$. 
Solving an \emph{assignment} problem of this bipartite graph is to find a matching with a maximum number of edges in which the sum of edge weights is as small as possible, which gives rise to an initial label assignment of leaves in $T^2$. 
In practice, the upper bound $\epsilon$ is chosen based on certain domain knowledge of the data; otherwise $\epsilon = \infty$ (which  is the case for all datasets in~\autoref{sec:results}). 

 \begin{figure}[ht!]
    \centering
    \includegraphics[width=1.02\columnwidth]{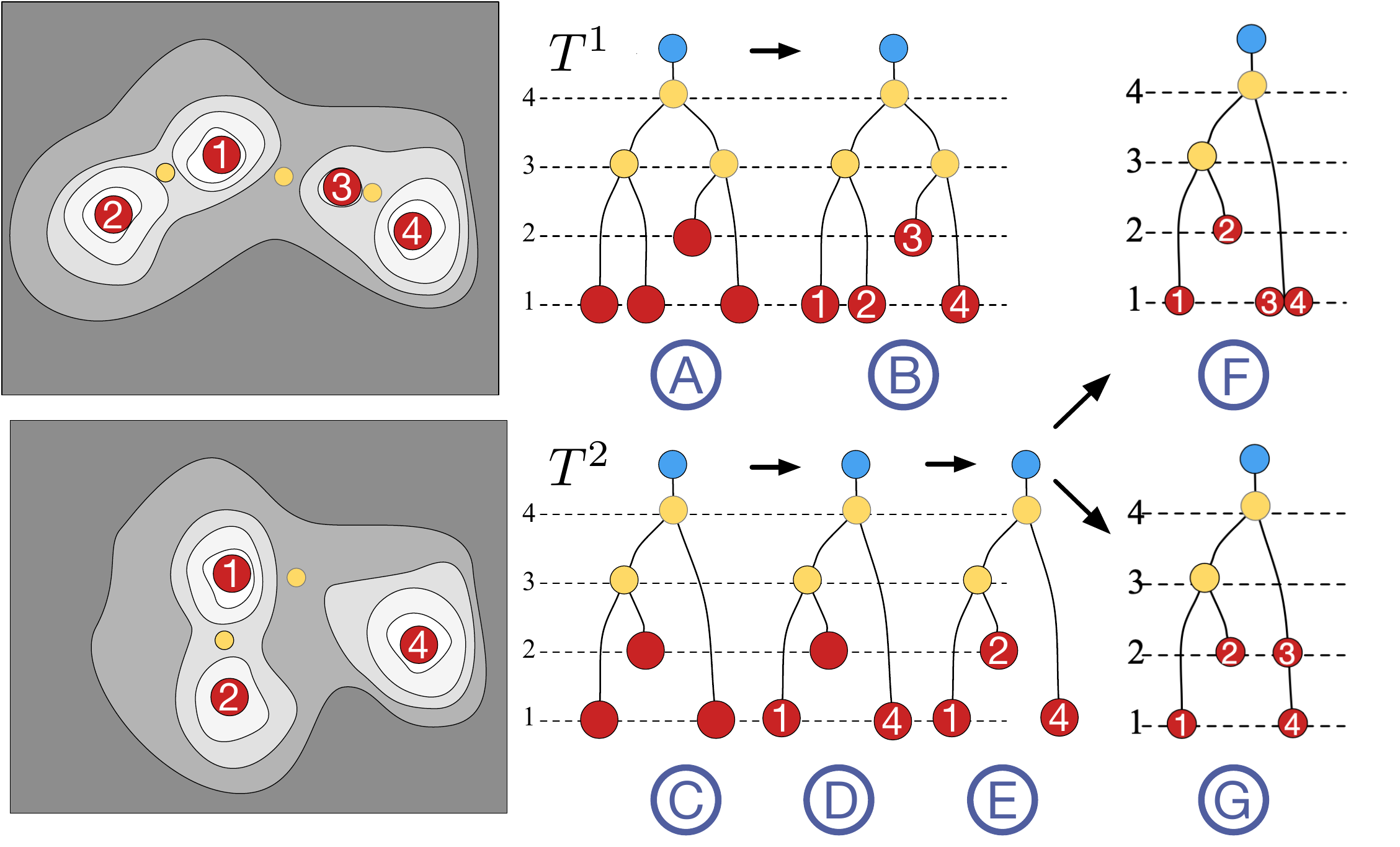}
    \vspace{-4mm}
    \caption{A labeling of leaves in $T^2$ (C) against a pivot tree $T^1$ (A) under the tree mapping strategy, using dummy leaves (F) or dummy vertices (G).}
\label{fig:tree-mapping}
\end{figure}

We illustrate this initial label assignment in~\autoref{fig:tree-mapping}. 
We start with a pair of merge trees $T^1$ and $T^2$ that arise from a pair of scalar fields. 
Since $|L(T^1)| = 4 > |L(T^2)| = 3$, $T^1$ in \autoref{fig:tree-mapping}A is chosen to be the pivot tree, which gives rise to a pivot label set $S_p=\{1,2,3,4\}$, see~\autoref{fig:tree-mapping}B.  
A subset of leaves in $T^2$ (\autoref{fig:tree-mapping}C) obtains their initial labels $\{1, 4\}$ by solving the above assignment problem (setting $\epsilon = 0.5$), since the respective leaves are close to one another; see \autoref{fig:tree-mapping}D.  
Thus, there are two unmatched labels $U_1 = \{2,3\}$ for $T^1$ and one unmatched label $U_2 = \{2\}$ for $T^2$.  

\begin{figure*}[!ht]
    \centering
    \includegraphics[width=1.0\textwidth]{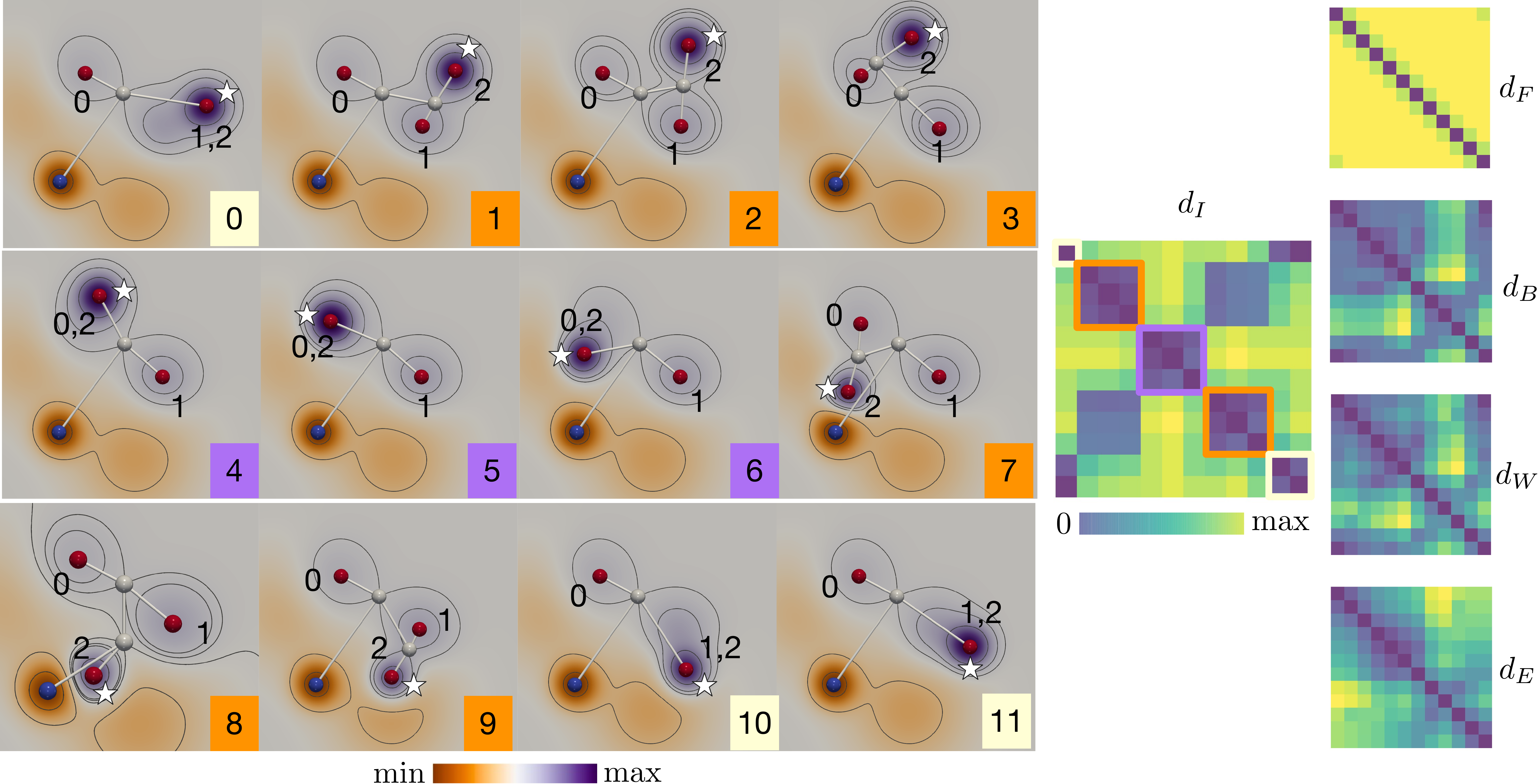}
    \vspace{-6mm}
    \caption{$\MG$ dataset with a hybrid mapping strategy and a time-varying pivot tree. Left: 12 time steps of scalar fields are visualized together with the labeling of the corresponding split trees. Middle and right: distance metrics for $d_F$, $d_B$, $d_W$, $d_E$, and $d_I$, respectively, where the zoomed-in version of $d_I$ highlights clusters among the data instances.} 
\label{fig:MG}
\end{figure*}

\para{Intermediate label assignment.}
We now construct a distances matrix $D^1$ between the unmatched labels and the matched labels in $T^1$ based on the hybrid  distance $d_h$. 
A matrix $D^2$ is constructed similarly. 
$D^1$ and $D^2$ by definition have the same number of columns. 
We now construct a complete, weighted bipartite graph between the unmatched labels of $T^1$ and $T^2$, where the weight $w_{x,y}$ between a label $x \in U_1$ and a label $y \in U_2$ is given by the $L_2$ distance between their corresponding rows in $D_1$ and $D_2$, respectively.
The final labeling of $L(T^2)$ is obtained by solving another  assignment problem in the above bipartite graph. 

We continue with the example in \autoref{fig:tree-mapping}.  
For simplicity, set $\lambda = 1$, $d_h$ is the same as the tree distance $d_t$.  
We obtain 
\[D_1 = 
\begin{pmatrix}
 4 & 6\\
 5 & 3
\end{pmatrix}, 
D_2 = 
\begin{pmatrix}
 3 & 5
\end{pmatrix}.
\] 
The weight between the label $2 \in U_1$ and label $2 \in U_2$ is 
$||\begin{pmatrix}
 4 & 6
\end{pmatrix}- \begin{pmatrix}
 3 & 5
\end{pmatrix}||_2 = 2$, and the weight between the label $3 \in U_1$ and label $2 \in U_2$ is 
$||\begin{pmatrix}
 5 & 3
\end{pmatrix}- \begin{pmatrix}
 3 & 5
\end{pmatrix}||_2 = 8$. 
By solving the above assignment problem, the unlabeled leaf in $T^2$ is assigned a label of 2; see~\autoref{fig:tree-mapping}E. 

\subsection{Dummy Leaves and Dummy Vertices}
\label{sec:dummy}

In general, after the intermediate label assignment, there may still be unmatched labels in the pivot tree; for example, the label $\{3\}$ remains unmatched for $T^2$ in~\autoref{fig:tree-mapping}E . 
The next step is to create dummy labels on $T^2$ to ensure that it uses the entire pivot label set. 
We introduce both \emph{dummy leaves} and \emph{dummy vertices} strategies. The former allows dummy labels only on the leaves, whereas the latter allows dummy labels on the interior of edges. 

\para{Dummy leaves.}
The dummy leaf strategy~\cite{YanWangMunch2020} finds an assignment for each unmatched label in the pivot tree $T^1$ by duplicating leaves in $T^2$. 
It uses a greedy assignment strategy based on the pairwise distance matrices.
The main idea is to find a leaf in $T^2$ that has the most similar local structure to an unmatched leaf in $T^1$. 
The $L_2$ distance between the rows from $D_1$ and $D_2$ serves as a similarity measure.

Using the example from~\autoref{fig:tree-mapping}F, $T^1$ has an  unmatched label 3. 
The $d_t$-based pairwise distance matrix between unmatched labels ($\{3\}$) and matched labels in $T^1$ $\{1,2,4\}$ is 
$D_1 = 
\begin{pmatrix}
 5 & 5 & 3
\end{pmatrix}
$. 
The $d_t$ based pairwise distance matrix of $T^2$ encodes distances between leaves $\{1, 2, 4\}$ and matched labels $\{1,2,4\}$, forming $D_2 = 
\begin{pmatrix}
 0 & 3 & 6\\
 3 & 0 & 5\\
 6 & 5 & 0
\end{pmatrix}
$. 
To find a leaf in $T^2$ that has the most similar local structure to the unmatched label 3 in $T^1$, rows from $D_1$ and $D_2$ are compared, which leads to a matching of leaf 3 in $T^1$ to leaf 4 in $T^2$. 
$T^2$ now contains one leaf with two labels $\{3, 4\}$, see \autoref{fig:tree-mapping}F. 

\para{Dummy vertices.}
The dummy leaf strategy is well-suited to generate smooth transitions between merge trees~\cite{YanWangMunch2020}, but it leads to instabilities in the distance computation. 
Therefore, we describe a second strategy that adds dummy vertices internal to the branches, 
which can also be interpreted as adding a branch with zero length. 
Thus, this strategy ensures that the dummy vertex does not change the tree structure of $T^2$. 
To determine the branch in $T^2$ where we add a dummy node, a pairwise distance matrix for all candidates is computed and compared to the distances in $T^1$.

Let us revisit the example in~\autoref{fig:tree-mapping}G. A dummy vertex with a label 3 is added to $T^2$ with the same scalar value as the leaf 3 of $T^1$. 
This dummy vertex has three candidates, resulting in a pairwise distance matrix  $D_2 =
\begin{pmatrix}
 1 & 2 & 5\\
 3 & 0 & 5\\
 5 & 4 & 1
\end{pmatrix}
$. Compared with $D_1 = 
\begin{pmatrix}
 5 & 5 & 3
\end{pmatrix}
$, the last candidate has the local structure most similar to leaf 3 of   $T^1$. The result is shown in~\autoref{fig:tree-mapping}G, where we add a dummy vertex on the branch of leaf 4 in $T^2$. 

\para{Induced matrices.}
After the labeling step, we transform a pair of unlabeled merge trees $T^1$ and $T^2$ to a pair of labeled merge trees $\T^1$ and $\T^2$.
We can transfer $\T^1$ and $\T^2$ to \emph{induced matrices} $\IM^1$ and $\IM^2$ according to label assignment, as shown in~\autoref{fig:tree-mapping}. 
The pivot label set gives rise to the rows and columns of an induced matrix, as shown below for the example in~\autoref{fig:tree-mapping}: 
 $$\IM^1=
\begin{pmatrix}
1 & 3 & 4 & 4\\
  & 1 & 4 & 4\\
  &   & 2 & 3\\
  &   &   & 1  
\end{pmatrix}, 
\IM^2=
\begin{pmatrix}
1 & 3 & 4 & 4\\
  & 2 & 4 & 4\\
  &   & 1 & 1\\
  &   &   & 1  
\end{pmatrix}.
$$

\subsection{Time-Varying Pivot Tree}
\label{sec:pivot-tree}

So far we have described strategies to find a leaf-leaf correspondence between a pair of unlabeled merge trees.
When moving to a time-varying dataset containing a large number of merge trees, this strategy is, however, not sufficient anymore.
It is desirable to have a shared label set for all trees resulting in comparable induced matrices with constant size.
Thus, the pivot tree selection plays an important role in our technique and we introduce different ways to approach this problem.

\para{Global pivot tree.}
The first method, first introduced in~\cite{YanWangMunch2020}, is a direct extension of the strategy from the matching of two trees described in~\autoref{sec:mapping} to many trees. 
It selects a \emph{global pivot tree} as a tree with the largest number of leaves among all input trees.
This tree defines the global label set, and we assign labels to all the other trees using the label set from the \emph{pivot tree}. 
A major limitation of this approach is the implicit assumption that similarity is a transitive property, which could lead to artifacts for large numbers of trees within a time-varying dataset.

\para{Time-varying pivot tree.}
To overcome this problem, we introduce a new, \emph{time-varying pivot tree} strategy. The labels are propagated from one tree to the next, capturing temporal changes in a time-varying dataset. 

The strategy works as follows: Given a set of merge trees $\{T^0, T^1, \dots, T^l\}$ that arises from a  time-varying dataset, 
let $T_p:=T^i$ (for some $i$) be an initial pivot tree with the largest $n$ number of leaves, thus defining the label set $[n]$. 
To assign labels to $T^{i-1}$ that immediately precedes $T^{i}$, we use $T^{i}$ as the pivot; thus, $T^{i-1}$ inherits labels from $T^i$. 
To assign labels to $T^{i-2}$, we use $T^{i-1}$ as its pivot instead. 
In general, $T^j$ will be the pivot tree for $T^{j-1}$ when $j \leq i$, whereas $T^j$ will be the pivot tree for $T^{j+1}$ when $j \geq i$. 
In a nutshell, the labels are inherited sequentially as we go through the dataset forward and backward from the initial pivot tree in a time-ordered way. 
Such a strategy works well with time-varying datasets, as demonstrated in~\autoref{sec:results}. 

\para{Pivot-free strategy.}
The time-varying pivot tree has its advantages and disadvantages.  
It is desirable for feature tracking within a time-varying dataset, thus supporting the detection of transitions and clusters in real-world datasets; see~\autoref{sec:transitions} and~\autoref{sec:clusters}. 
However, if the goal is to detect periodicities within time-varying datasets, we will need to effectively ``ignore'' geometric dependencies among adjacent time instances and treat these  instances independently, which leads to an alternative \emph{pivot-free} strategy.  
That is, we treat each time instance independently and compute  interleaving distances between pairs of instances without requiring a pivot tree or a shared label set across all input trees. 
In practice, this strategy works reasonably well if we assume the label sets are of roughly the same size; we give an example in~\autoref{sec:periodicities}.

\subsection{A Simple Example}
We end this section with a simple example involving a synthetic time-varying dataset, referred to as the $\MG$ dataset, to illustrate our analysis pipeline. 
This dataset is generated as a mixture of Gaussian functions centered at seven anchor points in a 2D domain. 
One of the anchor points (the starred point in~\autoref{fig:MG}A) performs a circular motion in the domain, while the rest remains stationary.  
This dataset contains 12 time steps modeled as scalar fields.  
We compute their corresponding split trees and pairwise distance matrices under $d_I$, $d_B$, $d_W$, and $d_E$, respectively (see \autoref{fig:MG}B). 
With the $\MG$ dataset, we demonstrate how geometric information coupled with topology helps to reveal its clustering structure, using hybrid mapping, dummy leaves, and time-varying pivot tree strategies. 
For all our experiments, unless otherwise specified, we set $\lambda = 0.5$ for a hybrid mapping strategy.

Geometric information and a time-varying pivot tree are the two key elements for tracking the moving Gaussian function. 
As shown in \autoref{fig:MG} left (time steps 0 to 11), using a hybrid mapping strategy, our method successfully tracks a moving Gaussian function centered at the starred critical point. 
This critical point with label 2 performs a counterclockwise circular motion: it merges with local maximum with labels $0$ and $1$ at time steps 4 and 10, respectively; and splits  with them at time steps 7 and 1, respectively. 
A hybrid mapping considers the geometric positions of critical points in the domain. Therefore, for adjacent instances, stationary critical points are more likely mapped with each other.

Furthermore, as shown in \autoref{fig:MG} right, in comparison with other distance metrics ($d_F$, $d_B$, $d_W$,  and $d_E$),  the labeled interleaving distance matrix detects three dominant clusters (bounded by purple, orange, and white squares). 
These clusters are the results of critical points merging and splitting in the time-varying data. 
In particular, the interactions of the local maximum 2 with other local maxima 0 and 1 at time steps 1, 4, 7, and 10 directly cause the topological transitions of the underlying split trees. 

On the other hand, we use a time-varying pivot tree strategy when we give labels to leaves. 
In this experiment, we pick $T^1$ as the initial pivot tree since $T^1$ has the largest number of leaves. Then $T^0$ and $T^2$ inherit labels from $T^1$. After that, $T^2$ becomes the pivot tree for $T^3$, and $T^3$ will inherit labels from $T^2$ and become the pivot tree for $T^4$, and so forth. If $T^p$ is the initial pivot tree, $T^i$ will be pivot tree for $T^{i+1}$ when $i \ge p$, whereas $T^i$ will be the pivot tree for $T^{i-1}$ when $i \leq p$. The benefit of using the time-varying pivot tree is that such a labeling strategy can propagate both geometric and topological information corresponding to temporal changes. If we used a global pivot tree strategy from~\cite{YanWangMunch2020}, $T^1$ is the only pivot tree, and current labeling results will change, especially when time instances are far from the global pivot tree temporally. For example, labeling $T^7$ and $T^8$ using $T^1$ as a pivot tree will be different with the current labeling result, no matter which mapping strategy we choose.

\begin{figure*}[!ht]
 \centering
  \includegraphics[width=1.0\linewidth]{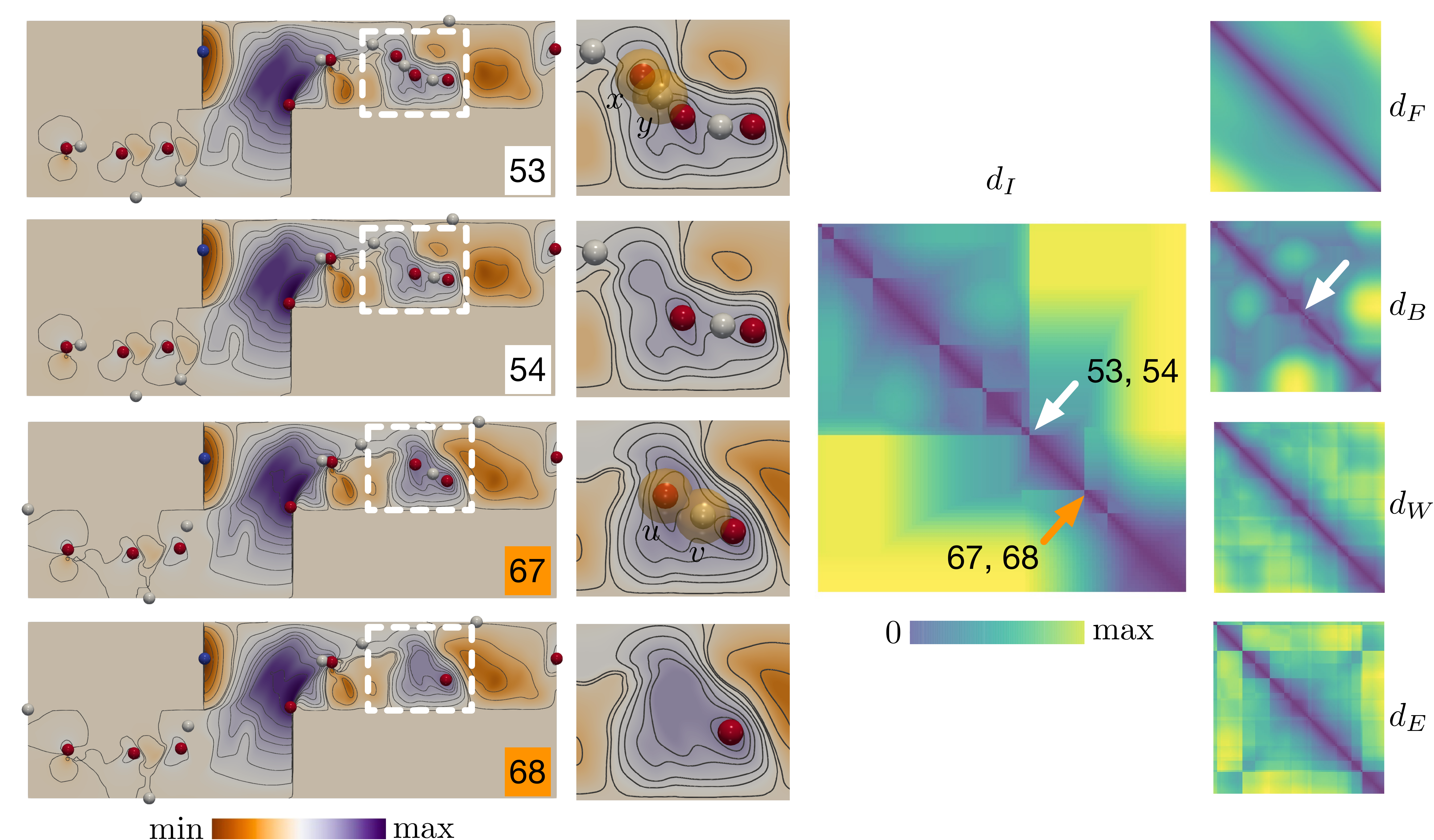}
  \vspace{-6mm}
  \caption{Geometry-aware merge tree comparisons for a time-varying {\CF} dataset. Using interleaving distances $d_I$, our framework helps to diagnose and highlight topological changes between adjacent time steps, in particular, between instances $53$ and $54$ and instances $67$ and $68$.}
  \label{fig:teaser}
  \end{figure*}
  
\begin{figure*}[!ht]
    \centering
    \includegraphics[width=\linewidth]{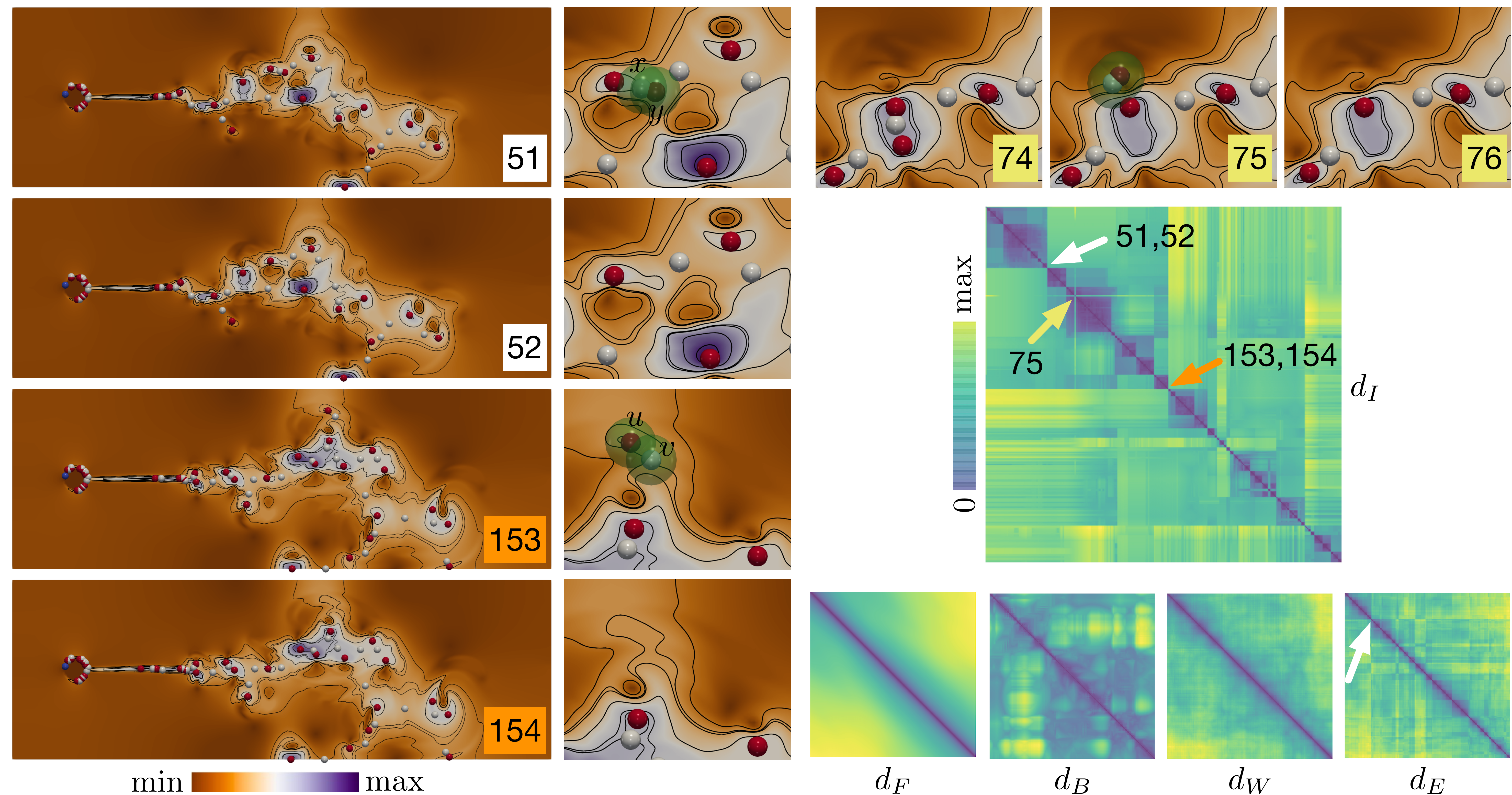}
    \vspace{-4mm}
    \caption{{\HF} dataset: geometry-aware merge tree comparisons. Left two columns: selected scalar fields with their zoomed-in views, where local minimum are in blue, saddles are in white, local maximum are in red. Right two columns: pairwise distance matrices for $d_F$, $d_B$, $d_W$,  $d_E$, and $d_I$, respectively. Arrows in $d_I$ highlight detected structural transitions.} 
\label{fig:HC}
\end{figure*}

\section{Experimental Results}
\label{sec:results}

In this section, we demonstrate via experiments that geometry-aware merge tree comparisons based on the interleaving distance help to detect \emph{transitions}, \emph{clusters}, and \emph{periodicities} of a time-varying dataset, as well as to \emph{diagnose} and \emph{highlight} the topological changes between adjacent instances.

\subsection{Detect and Diagnose Structural Transitions}
\label{sec:transitions}

We demonstrate our method in detecting structural transitions using two flow datasets, namely, the $\CF$ and $\HF$ datasets.  

\begin{figure}
    \centering
    \includegraphics[width=1.0\columnwidth]{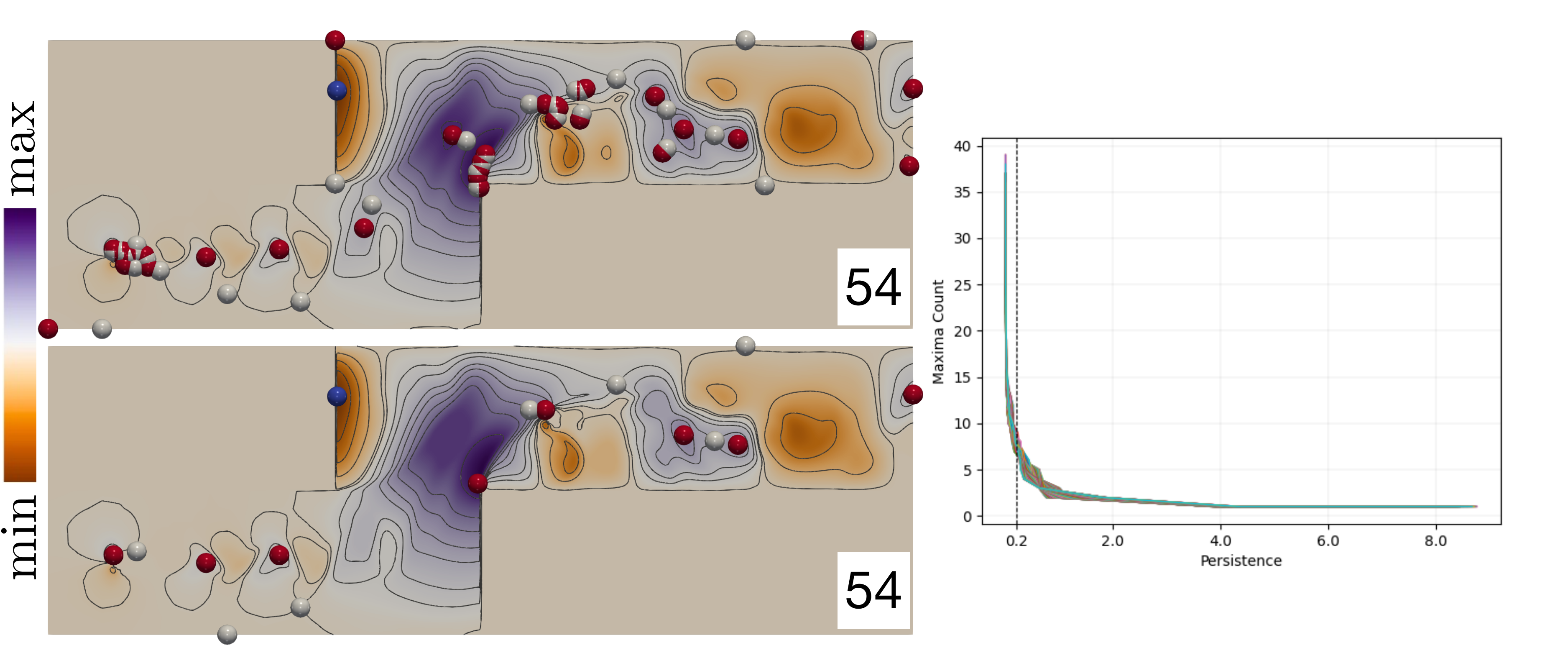}
    \vspace{-6mm}
    \caption{$\CF$ dataset. Left: time instance 54 before (top) and after (bottom) persistence simplification. Right: a set of persistence graphs used to guide the simplification process.} 
\label{fig:corner-flow-simplify}
\end{figure}

\para{Corner Flow dataset.} 
We first demonstrate our method using the 2D Cylinder Flow Around Corners dataset\footnote{\label{TDA-DATA}https://cgl.ethz.ch/research/visualization/data.php}, which we refer to as the $\CF$ dataset. 
This dataset arises from the simulation of a viscous 2D flow around two cylinders~\cite{BaezaRojoGunther2020, Popinet2004}. The channel into which the fluid is injected is bounded by solid walls. 
A vortex street is initially formed at the lower left corner, which then evolves around the two corners of the bounding walls. 
We generate a set of split trees from the vertical component of the velocity vector fields based on 94 time instances -- they correspond to steps 801-894 from the original 1500 time steps. 
These instances describe the formation of a one-sided vortex street on the upper right corner; see Fig.~\ref{fig:teaser} left that visualizes the scalar fields associated with time steps 53, 54, 67, and 68.

To separate signals from noise, we apply persistence simplification~\cite{EdelsbrunnerLetscherZomorodian2002} to the split trees of all instances with a persistence threshold of $0.2$. 
In particular, to guide the selection of the persistence threshold, we employ a set of \emph{persistence graphs}, each of which represent the number of persistence pairs as a function of persistence~\cite{GerberBremerPascucci2010}.  
The shape of the persistence graph, in particular, a  plateau,  indicates  a  stable  range  of scales  to  separate  noise  from  signals in the persistence graph~\cite{BremerMaljovecSaha2015, AthawaleMaljovecYan2020}.
We demonstrate such a simplification process in \autoref{fig:corner-flow-simplify}, where time instance 54 is shown before and after persistence simplification; the persistence threshold is chosen approximately due to the variabilities across time instances.

Our framework analyzes structural transitions via the pairwise distance matrices. For this experiment, we use hybrid mapping, dummy vertex,  and time-varying pivot tree strategies. As shown in~\autoref{fig:teaser} middle, our framework using the interleaving distance $d_I$ captures two obvious structural transitions among adjacent instances, $53 \to 54$ and $67 \to 68$. 
There appear to be clear block structures within the $d_I$ matrix, where the above transitions are highlighted by arrows at the corners of these blocks. 

Under the diagnostic setting, we locate critical points in the domain that are responsible for the $d_I$ distance between adjacent instances. 
A close inspection of this time-varying dataset then reveals that from step 53 to 54, a pair of critical points $x$ and $y$ (enclosed by orange spheres) disappears. 
Similarly, from step 67 to 68, another pair of critical points $u$ and $v$ (enclosed by orange spheres) disappears.  
Therefore, $d_I$ highlights structural transitions in the time-varying data, whereas in comparison, only the bottleneck distance $d_B$ is able to capture the same $53 \to 54$ transition in its matrix representation (\autoref{fig:teaser} white arrow in $d_B$). 

\para{Heated Flow dataset.}
We give another example using a 2D Heated Cylinder with Boussinesq Approximation dataset\footref{TDA-DATA}, denoted as the {\HF} dataset. 
This dataset comes from the simulation of a 2D flow generated by a heated cylinder using the Boussinesq approximation~\cite{GuntherGrossTheisel2017, Popinet2004}. 
It shows a time-varying turbulent plume containing numerous small vortices.  

We convert each time instance of the flow into a scalar field using the magnitude of the velocity vector. 
We then generate a set of split trees from these scalar fields based on 300 time steps -- they correspond to steps 1000-1299 from the original 2000 time steps. 
This dataset captures the evolution of small vortices over time. 
We use hybrid mapping, dummy vertex, and time-varying pivot tree strategies, and apply simplification with a persistence threshold of $0.06$. 

The results are shown in~\autoref{fig:HC}. 
We observe two visible structural transitions based on $d_I$ between steps $51 \to 52$ and $153 \to 154$ (pointed by white and orange arrows, respectively).  
Under the diagnostic setting, the structural transition $51 \to 52$ is caused by the disappearance of a pair of critical points $x$ and $y$ at step 51 (highlighted by green bubbles). 
The transition $153 \to 154$ is a result of the disappearance of the pair $u$ and $v$ at step $153$. 

Two additional structural transitions are detected at $74 \to 75$, and  $75 \to 76$, as indicated by yellow arrow in the $d_I$ matrix of~\autoref{fig:HC}. 
However, a closer inspection indicates that such a transition is, in fact, an artifact as a result of persistence simplification, which leads to structural changes of the simplified split trees.   
We will discuss such artifacts further in~\autoref{sec:discussion}. 
 
\begin{figure*}[ht]
    \centering
    \includegraphics[width=\linewidth]{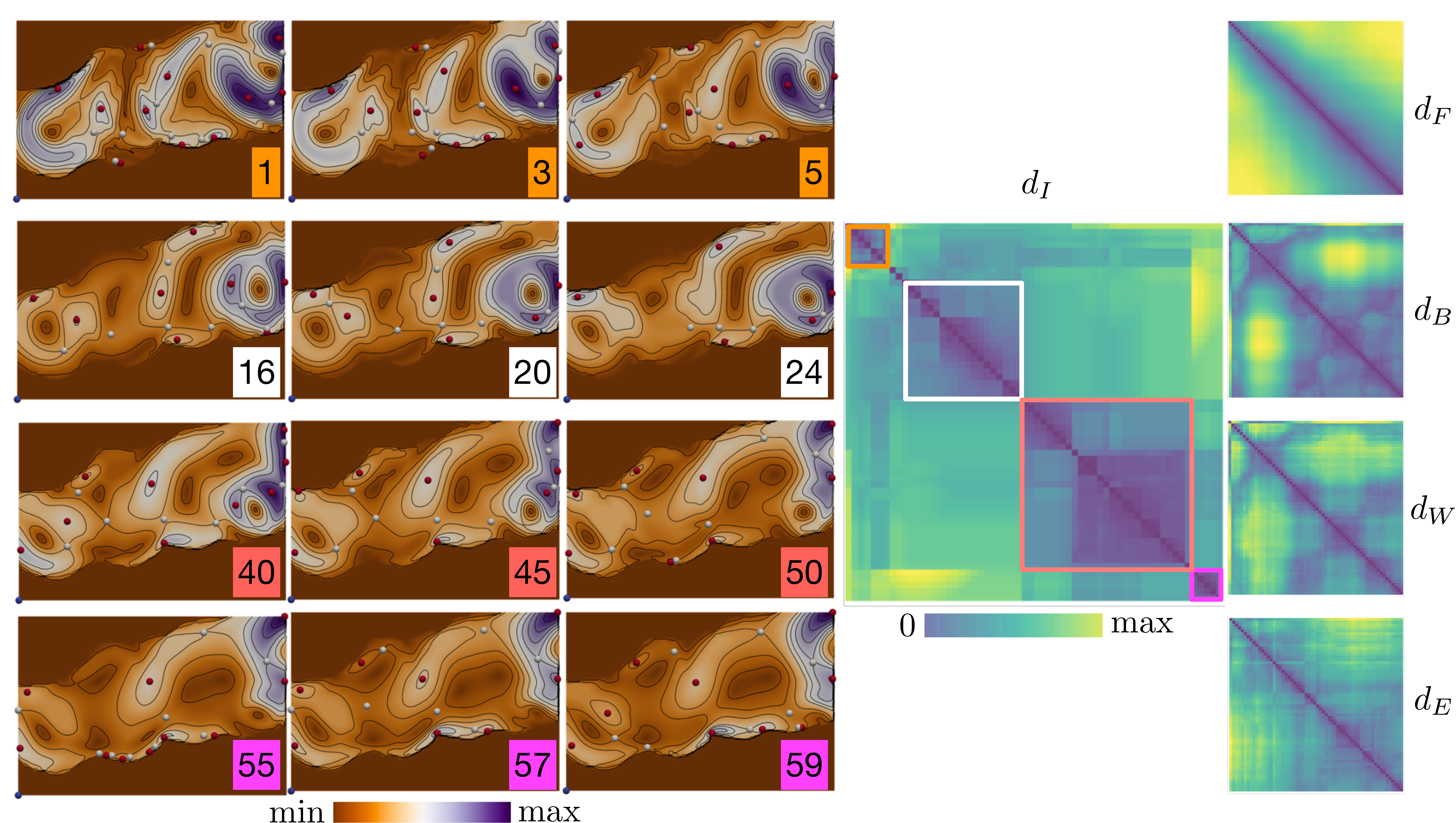}
    \vspace{-4mm}
    \caption{{\RS} dataset: detect clusters. Left three columns: selected scalar fields drawn from each of the four clusters. Right two columns: pairwise distance matrices for $d_F$, $d_B$, $d_W$, $d_E$, and $d_I$, respectively. Colored boxes in $d_I$ highlight the four detected clusters.} 
\label{fig:RS}
\end{figure*}

\begin{figure*}[ht]
    \centering
    \includegraphics[width=\linewidth]{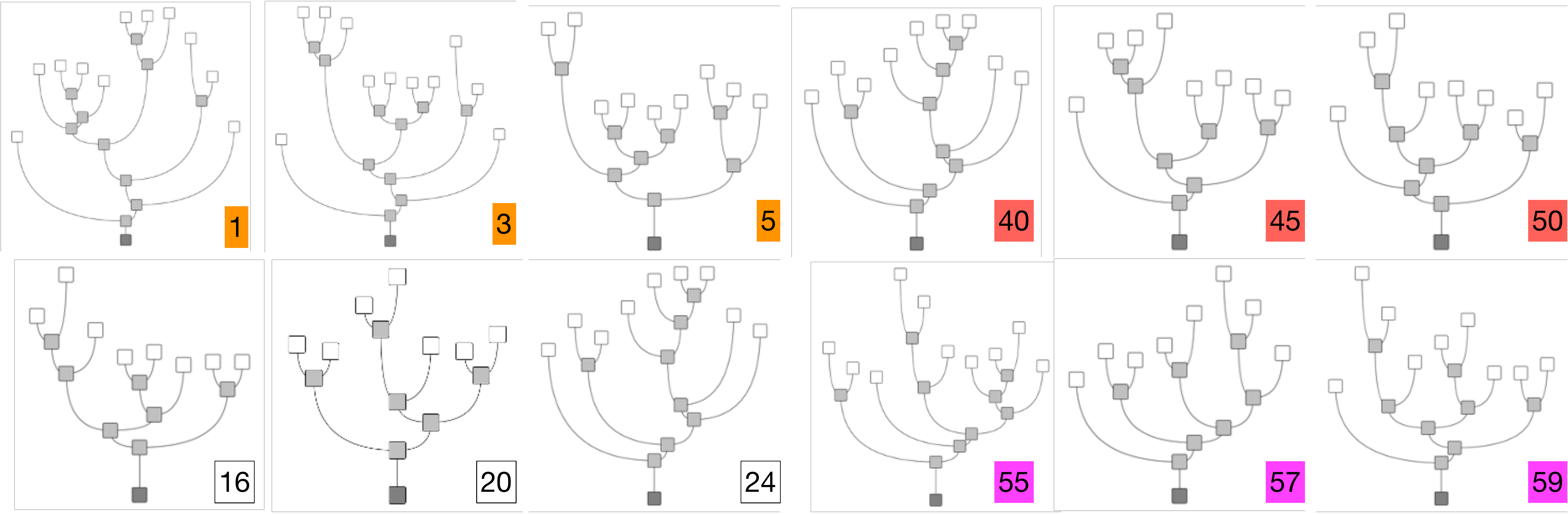}
    \vspace{-4mm}
    \caption{{\RS} dataset: abstract split trees for selected time steps, {\cf}~\autoref{fig:RS}.} 
\label{fig:RS-ST}
\end{figure*}

\subsection{Detect Clusters}
\label{sec:clusters}

Our method using the interleaving distance helps to cluster time instances based on their structural differences. 
We demonstrate the utility of the method using 2D simulations of the Red Sea and a dataset generated from Gerris flow solver. 

\para{Red Sea dataset.}
The {\RS} dataset originates from the IEEE Scientific Visualization  Contest 2020\footnote{https://kaust-vislab.github.io/SciVis2020/}, and is generated using a high-resolution MITgcm  (Massachusetts Institute of Technology general circulation model), together with remote sensing satellite observations.
It is used to study the circulation dynamics and eddy activities of the Red Sea  (see~\cite{HoteitLuoBocquet2018, ZhanKrokosGuo2019, ZhanSubramanianYao2014}). 
For the experiment, we use the velocity magnitude fields of a particular dataset (named \emph{001.tgz}) with 60 time steps. 
We generate split trees from the 2D slices perpendicular to the z-axis ($z=1$).
For this experiment, we use hybrid mapping, dummy vertex, and time-varying pivot tree strategies.

\begin{figure*}[ht]
    \centering
    \includegraphics[width=\linewidth]{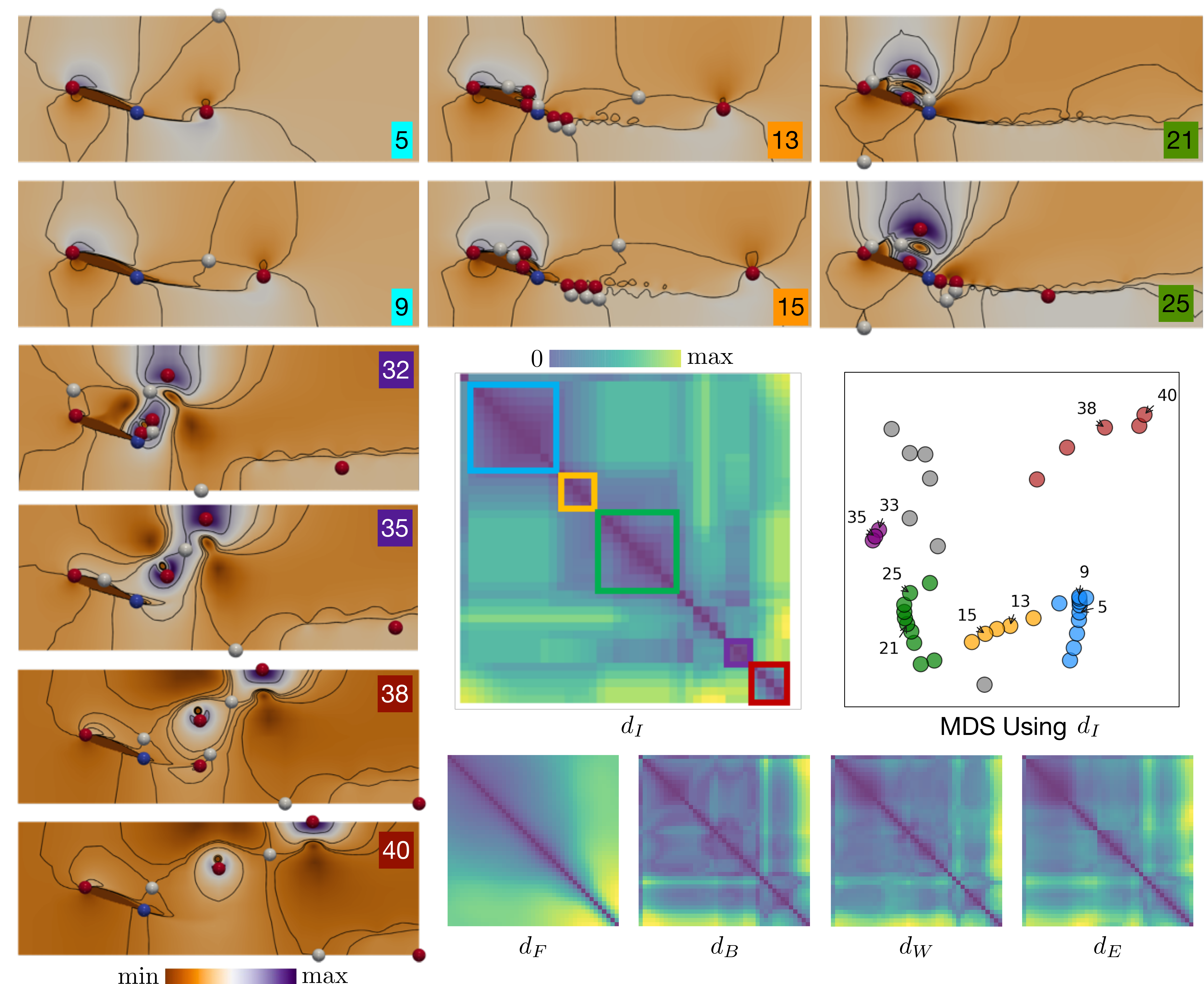}
    \vspace{-4mm}
    \caption{{\Wing} dataset: detect clusters. Left column and top two rows: selected scalar fields drawn from each of the five clusters. 
      Right column and third row: MDS projection of all the time steps using the $d_I$ metric. 
      Res: pairwise distance matrices for $d_F$, $d_B$, $d_W$, $d_E$, and $d_I$, respectively. Colored boxes in $d_I$ highlight the five detected clusters.} 
\label{fig:wing}
\end{figure*}

As shown in~\autoref{fig:RS}, the $d_I$ distance matrix clearly indicates four clusters of data instances. 
It shows that the scalar fields share similar structures from instances 0-6 (in orange square), 10-27 (in white square), 28-54 (in red square), and 55-59 (in magenta square). 
We visualize several selected scalar fields in~\autoref{fig:RS} left. 

Taking a closer look at their corresponding split trees in~\autoref{fig:RS-ST}, each cluster of data instances share a similar tree structure, especially for some instances in the red cluster (e.g., see instances 40, 45, and 50). 
In this experiment, $d_B$ and $d_W$ also show some clustering patterns; however, in comparison, $d_I$ offers clearer and more informative clustering pattern.

\para{Wing dataset.}
The {\Wing} dataset is generated using the software \textit{Gerris flow solver} \footnote{http://gfs.sourceforge.net/}. 
We use its demo flow simulation example involving the ``starting vortex'', which is a vortex that forms in the fluid near the trailing edge of an aerofoil (wing) as it is accelerated from rest in a fluid. 
For our simulation, we set the angle of the aerofoil with respect to the fluid as $20^{\circ}$ and generate $40$ time steps. The scalar field of interest is the velocity magnitude. 

For this experiment, we use hybrid mapping, dummy vertex, and time-varying pivot tree strategies. 
As shown in~\autoref{fig:wing}, we compute various distance matrices based on the split trees computed for the velocity magnitude field. All the trees are simplified at a simplification threshold of $0.13$.

As shown in~\autoref{fig:wing}, the $d_I$ matrix clearly detects five clusters of data instances. 
The MDS projection of all data instances using $d_I$ further highlights the clustering structures among the scalar fields at different time steps. 
Whereas $d_B$, $d_W$, and $d_E$ also capture some clustering structures in this time-varying dataset, $d_I$ gives rise to clusters that appear to be more separable and visually differentiable. Moreover, based on inspection of the velocity magnitude fields shown in ~\autoref{fig:wing}, it is apparent that the five clusters separate the time steps into clusters with similar vortex structures. For example, the major change between the orange and green clusters is the appearance of a strong vortex on the top of the wing in the green cluster. Similarly, notice the two strong vortices in the time steps corresponding to the purple cluster. As one of those vortices exits the domain in the subsequent time steps, the instances are grouped into a different cluster marked as red.  

\begin{figure*}[h!]
    \centering
    \includegraphics[width=\linewidth]{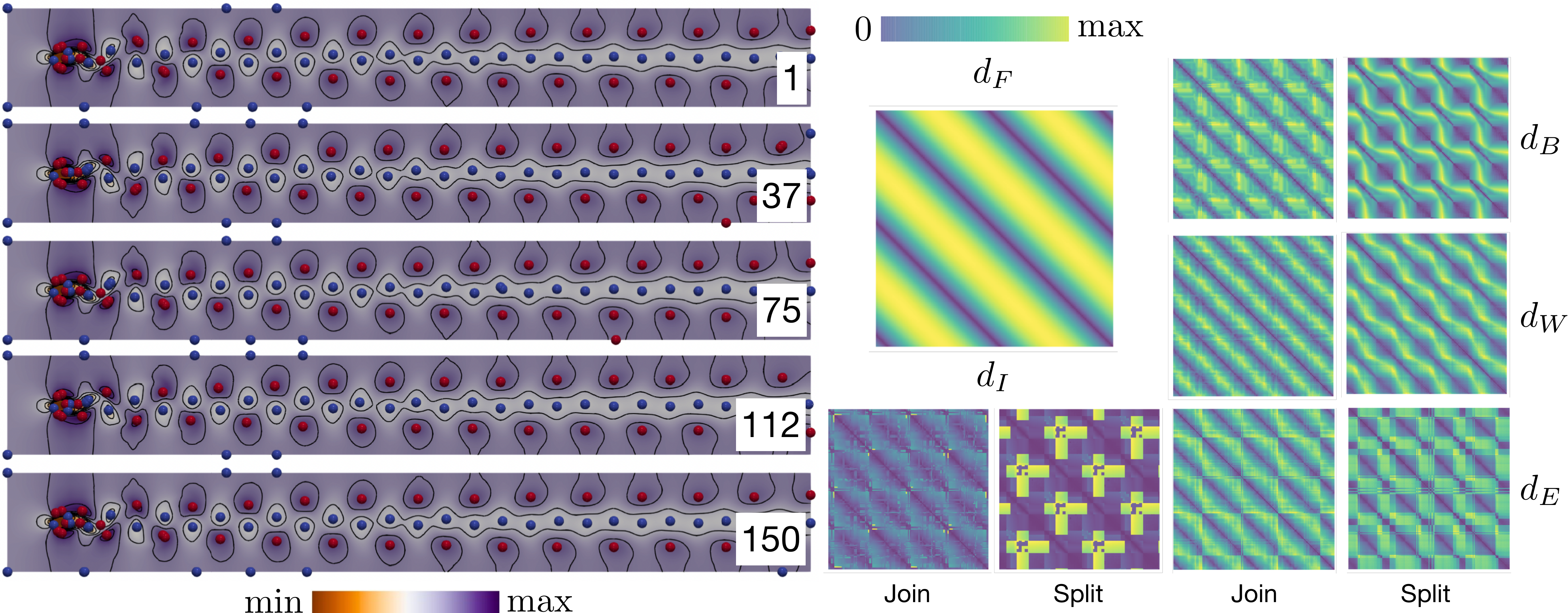}
    \vspace{-4mm}
    \caption{$\mathsf{Vortex Street}$ dataset. Left: five selected time steps are shown for the dataset; notice that the scalar field is mirrored after 37 steps and goes back to the original after 75 time steps.  Middle: distance matrices for $d_F$ and $d_I$, respectively; Right: distance matrices for $d_B$, $d_W$, and $d_E$, using both join and split trees, respectively.}
    \label{fig:VS}
\end{figure*}
 
\subsection{Detect Periodicities}
\label{sec:periodicities}

Finally, we demonstrate our framework in detecting periodicities using the classic 2D von K\'arman vortex street dataset, which we refer to as the {\VS} dataset. 
We consider the region of vortex shedding behind the cylinder, and use the velocity magnitude field for comparison as used earlier by Sridharamurthy \etal~\cite{SridharamurthyMasoodKamakshidasan2020}.  

We use hybrid mapping, dummy vertices, and pivot-free strategies. 
As shown in~\autoref{fig:VS}, using either the join or split tree, $d_B$, $d_W$, and $d_E$ all show a periodicity of length $37$. 
The interleaving distance, $d_I$, captures the same length of periodicity using the join tree.  
However, using the split tree, we detect a periodicity of length 75 using $d_I$. 
Such a longer periodicity coincides with the periodicity detected using $d_F$. 
This periodicity can be justified where both $d_I$ and $d_F$ consider more geometric information in the domain, in comparison with other metrics. 

Furthermore, as shown in~\autoref{fig:VS} left, the positions of local maxima (red points) change more drastically every 37 time steps, in comparison with the positions of local minima (blue points). 
This finer difference is captured by the split tree version of $d_I$; where neither $d_B$, $d_W$, nor $d_E$ capture this difference in the behavior of the local extrema.

In this experiment, as mentioned in~\autoref{sec:method}, a pivot-free mapping strategy works better than a time-varying pivot tree strategy in detecting periodicities.

\section{Conclusion and Discussion}
\label{sec:discussion}

In this paper, we introduce a systematic way to integrate geometric  information for comparing merge trees. 
Given a pair of merge trees that arise from scalar fields, our main idea is to decouple the computation of a distance measure into two steps: a labeling step that generates a correspondence between critical points of the merge trees, and a distance computation step that computes the labeled interleaving distance between a pair of labeled merge trees by encoding them as matrices. 
To encode geometric information, we introduce a hybrid strategy during the labeling step that considers the intrinsic tree distances between critical points and/or the Euclidean distances between  their locations in the data domain. 
We demonstrate that our approach can be used to detect clusters, structural transitions, and periodicity in a way that is either comparable or complementary to existing approaches. 
There are many directions for future research. 

\para{Improved efficiency and robustness.} 
Naively computing the labeled interleaving distance $d_I$ requires access to all entries in the reduced matrix, which takes $O(n^2)$ time. 
Look into a more scalable computation would be interesting, possibly taking inspiration from the work of Kerber \etal~\cite{KerberMorozovNigmetov2017}.  

In addition, the labeled interleaving distance coupled with various labeling strategies has strengths and weaknesses. 
The strategy is shown to detect periodicity that is more sensitive to the underlying geometry, and it enjoys a certain amount of robustness due to its stability properties.  
However, at the same time, it is not as robust as some other metrics  when the underlying data contain a large amount of noise and a large number of features; improving its robustness is left for future work. 

\para{Integration of domain knowledge.} Our two-step comparative process opens doors to the integration of domain knowledge. 
The labeling process can be easily extended to not only integrate geometric information from the data domain but also to encode information from its underlying applications.  
In particular, domain knowledge will be useful during the initial label assignment described in~\autoref{sec:mapping}. 

\begin{figure}[!ht]
    \centering
    \includegraphics[width=.99\columnwidth]{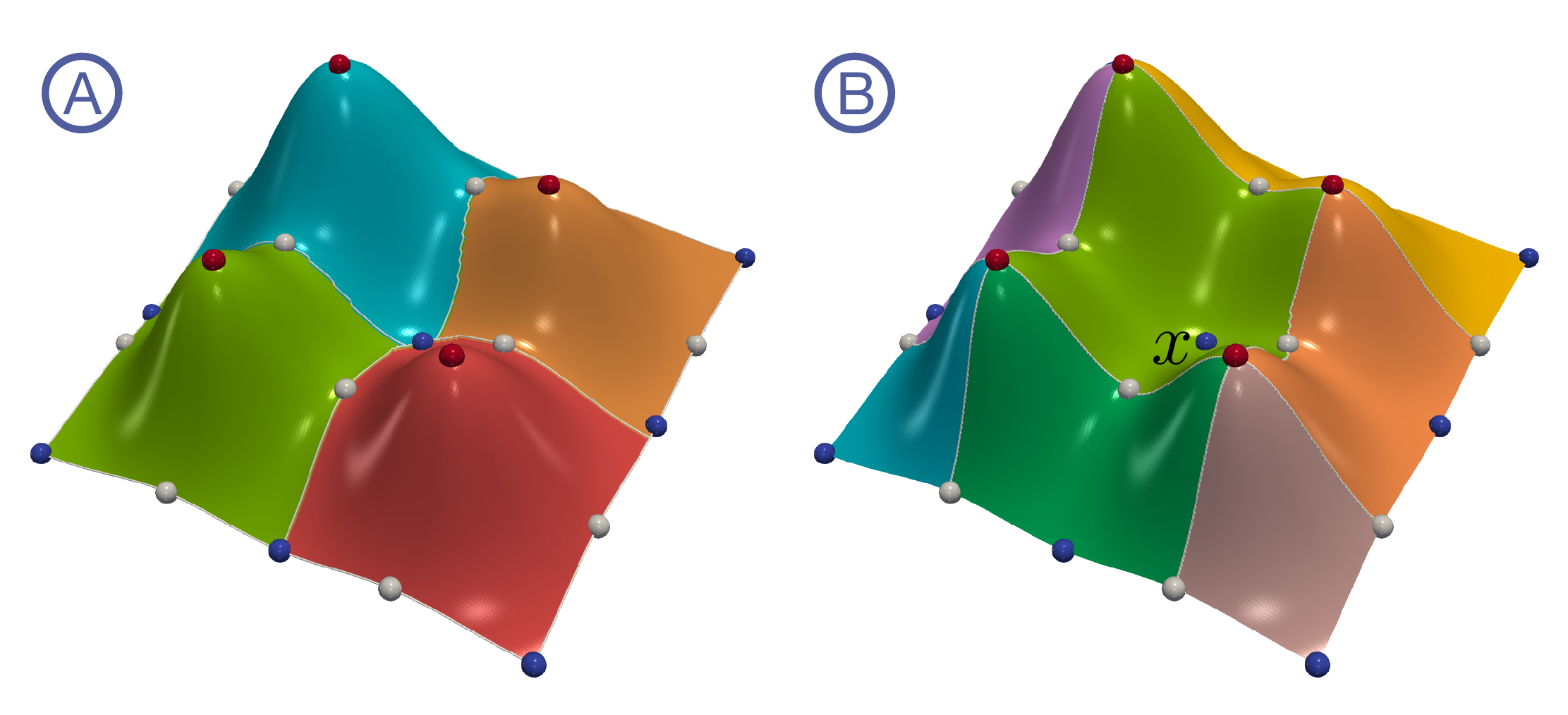}
    \vspace{-2mm}
    \caption{Given the 2D function $f$. (a) shows the stable manifolds forming the Morse complex of $f$. (b) shows the unstable manifolds forming the Morse complex of $-f$; the green cell surrounding the critical point $p$ is an unstable manifold of $p$.} 
\label{fig:morse-complex}
\end{figure}

\para{Other geometry-based labeling strategies.} 
The hybrid labeling strategy based on the tree distance and/or the Euclidean distance between critical points is just one example among many possible geometry-aware labeling strategies. 
For example, we could adapt a strategy called the \emph{Morse mapping} that has been used for the tracking of critical points~\cite{ReininghausKastenWeinkauf2012}. 
Given a pair of merge trees $T^1$ and $T^2$, the Morse mapping strategy facilitates the gradient flow derived from the scalar fields to define a forward $T^1 \rightarrow T^2$ and a backward $T^2 \rightarrow T^1$ mapping.
The forward and backward assignment builds on the partition of the domain provided by the Morse complex~\cite{EdelsbrunnerHarerZomorodian2001}, which represents the gradient behavior of the scalar field of the data domain. 
 See~\autoref{fig:morse-complex} for an example of a Morse complex of a 2D scalar field. 

Let $\grad{f}$ denote the gradient of a Morse function $f: \Xspace \to \Rspace$.  
An \emph{integral line} at a regular point is a maximal path whose tangent vectors agree with the gradient \cite{EdelsbrunnerHarerZomorodian2001}. 
The function increases along an integral line, which begins and ends at critical points.   
The \emph{stable manifold} (or \emph{unstable manifold}) surrounding a critical point $x$ includes $x$ itself and all regular points whose integral lines end (or originate) at $x$~\cite[Chap.~VI, page 131]{EdelsbrunnerHarer2010}. 
The stable and unstable manifolds of $x$ are denoted as $S(x)$ and $U(x)$, respectively. 
To define the Morse mapping strategy, we use the unstable manifolds surrounding the local minima of the scalar field (\autoref{fig:morse-complex}b), which correspond to leaves in the merge tree.

\begin{figure}[!ht]
    \centering
    \includegraphics[width=.99\columnwidth]{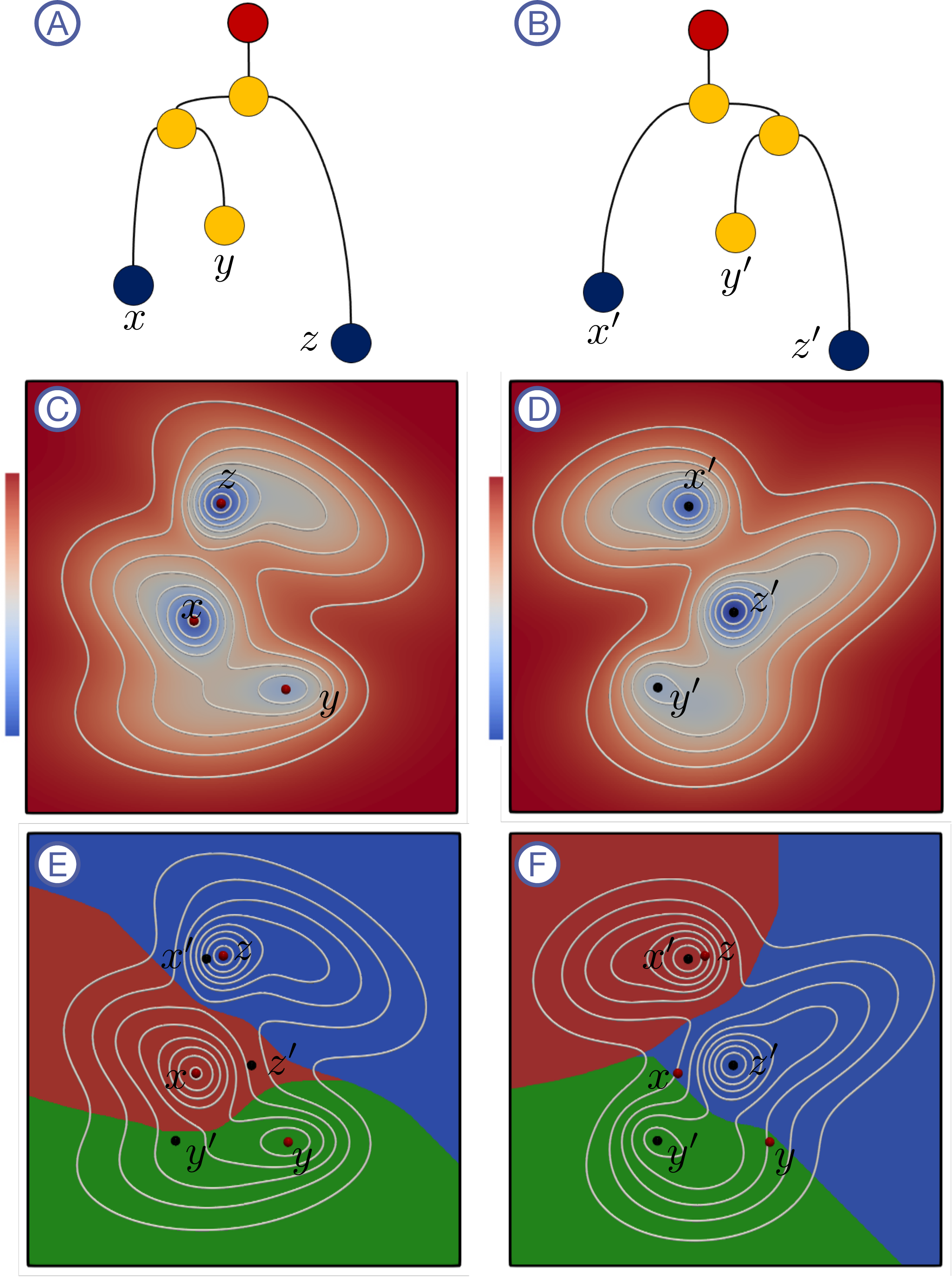}
    \vspace{-2mm}
    \caption{Comparing three mapping strategies with two synthetic datasets generated as mixtures of Gaussians. Label assignments obtained via tree mapping, Euclidean mapping, and Morse mapping strategies give rise to different labels.} 
\label{fig:compare-mapping}
\end{figure}

Suppose we are given a pair of merge trees $T^1$ and $T^2$ that arise from a pair of scalar fields.
Let $x\in V(T^1)$ and $x'\in V(T^2)$ denote a pair of leaves (local minima of the underlying scalar fields). 
Let $U(x)$, $U(x')$ denote their respective unstable manifolds.
We say $x$ is \emph{forward mapped} to $x'$ if $x \in U(x')$ and $x'$ is \emph{backward mapped} to $x$ if $x' \in U(x)$, denoted as $x \rightarrow x'$ and $x \leftarrow x'$, respectively. 
In other words, we check to which unstable manifold a leave belongs to determine the label assignment. 
As illustrated in~\autoref{fig:compare-mapping}E-F, given a pair of merge trees $T^1$ and $T^2$ (\autoref{fig:compare-mapping}A-B) that arise from scalar fields (\autoref{fig:compare-mapping}C-D), local minimum $x$ of $T^1$ is forward mapped to $z'$ since $x \in U(z')$, whereas $z'$ is backward mapped to $x$ since $z' \in U(x)$. 

This strategy leads to three categories of matched leaf pairs: \emph{double connected pairs}, which result in a joint label; and  \emph{forward} and \emph{backward connected pairs},  which generate a novel label and introduce a dummy node in one of the trees. 
For example, $x$ and $z'$ forms a double connected pair in~\autoref{fig:compare-mapping}.  

Finally, \autoref{fig:compare-mapping} further illustrates that different mapping strategies between two merge trees result in different label assignments.  
In this example, merge trees $T^1$ and $T^2$ arise from two synthetic 2D scalar fields generated as mixtures of Gaussians. 
$T^1$ (\autoref{fig:compare-mapping}A) contains three leaves that correspond to local minima $x, y, z$ in the domain (\autoref{fig:compare-mapping}C).  
$T^2$ (\autoref{fig:compare-mapping}B) contains three leaves that correspond to local minima $x', y', z'$ (\autoref{fig:compare-mapping}C).  
Using the tree mapping strategy, we obtain a bijective mapping $x \leftrightarrow x'$, $y \leftrightarrow y'$, and $z \leftrightarrow z'$ (\cf, \autoref{fig:compare-mapping}A-B). 
Using the Euclidean mapping strategy, we obtain a different bijective mapping,  $x \leftrightarrow y'$, $y \leftrightarrow z'$, and $z \leftrightarrow x'$ (\cf, \autoref{fig:compare-mapping}C-D). 
Finally, using the Morse mapping, we obtain sets of forward ($x \rightarrow z'$, $y \rightarrow y'$, and $z \rightarrow x'$, \autoref{fig:compare-mapping}F) and backward ($x \leftarrow z'$, $y \leftarrow y'$, and $z \leftarrow x'$, \autoref{fig:compare-mapping}E) mappings, forming double connected pairs.    
Understanding such differences is important in choosing the appropriate strategies for particular datasets, which remains an open question.  

\section*{Acknowledgments}
This project was partially supported by NSF IIS 1910733 and 
DOE DE-SC0021015. 


\vskip -2.2\baselineskip plus -1fil
\begin{IEEEbiographynophoto}{Lin Yan}
is a PhD student in the Scientific Computing and Imaging (SCI) Institute, University of Utah. Her research interests include  topological  data  analysis  and  visualization.  Her  recent  work  includes statistical analysis and uncertainty visualization of topological descriptors. 
\end{IEEEbiographynophoto}
\vskip -2.2\baselineskip plus -1fil
\begin{IEEEbiographynophoto}{Talha Bin Masood}
is a Postdoctoral Fellow at Link\"{o}ping University in Sweden. 
He received his Ph.D. in Computer Science from the Indian  Institute  of  Science,  Bangalore.  His  research  interests  include  scientific  visualization,  computational  geometry,  computational topology, and their applications to various scientific domains.
\end{IEEEbiographynophoto}
\vskip -2.2\baselineskip plus -1fil
\begin{IEEEbiographynophoto}{Farhan Rasheed}
is PhD student at Link\"{o}ping University in Sweden. He graduated from Heidelberg University with a degree in Scientific Computing. His research interests includes scientific visualization, topological data analysis, machine learning, and medical image computing.
\end{IEEEbiographynophoto}
\vskip -2.2\baselineskip plus -1fil
\begin{IEEEbiographynophoto}{Ingrid Hotz}
is currently a Professor in Scientific Visualization at the Link\"{o}ping University in Sweden. She received her Ph.D. degree from the Computer Science Department at the University of Kaiserslautern, Germany. Her research interests lie in data analysis  and  scientific  visualization,  ranging  from  basic  research questions  to  effective  solutions  to  visualization  problems  in  applications.  
\end{IEEEbiographynophoto}
\vskip -2.2\baselineskip plus -1fil
\begin{IEEEbiographynophoto}{Bei Wang}
is  an  Assistant  Professor  at  the  School  of  Computing and a faculty member at the SCI Institute,  University  of  Utah.  She  received  her  Ph.D.in  Computer  Science  from  Duke  University.   Her research  interests  include  topological  data  analysis,  data  visualization,  computational  topology,  computational  geometry,  machine learning, and data mining. 
\end{IEEEbiographynophoto}

\end{document}